\documentclass[]{spie}  
 
\usepackage{amsmath,amsfonts,amssymb}
\usepackage{graphicx}
\usepackage[colorlinks=true, allcolors=blue]{hyperref}
\usepackage{multirow}
\usepackage{gensymb}
\usepackage{subfig}
\usepackage{multicol}
\usepackage[table,xcdraw]{xcolor}
\usepackage{lineno}

\title{LUVCam: A high-performance, low-cost, UV/optical camera for the future of astronomy in space}

\author[a,f,h]{Aaron Tohuvavohu}
\author[a,b]{Suresh Sivanandam}
\author[a,c]{Jean-Christophe Fronteddu}
\author[a]{Patrick Nkwari}
\author[a]{Gavin Hay}
\author[a]{Shaojie Chen}
\author[a]{Sarik Jeram}
\author[a]{Albert Lau}
\author[a]{Jacob Taylor}
\author[a]{Mark Barnet}
\author[e]{Andras Pal}
\author[a]{Mohamed Shaaban}
\author[g]{Ajay S. Gill}
\author[a]{Julia Empey}
\author[a]{Emma Seabrook}
\author[a]{Coby Silayan}
\author[a]{Dhwanil Patel}
\author[a]{Braden Seefeldt-Gail}
\author[a,b]{Phil R. Van-Lane}
\author[a]{Scott Christie}
\author[b]{Christopher D. Matzner}
\author[c]{Chris Damaren}
\author[b]{Maria Drout}
\author[d]{Filip Munz}
\author[d]{Jakub \v{R}\'{i}pa}
\author[d]{Norbert Werner}

\affil[a]{Dunlap Institute for Astronomy \& Astrophysics, University of Toronto}
\affil[b]{David A. Dunlap Department of Astronomy \& Astrophysics, University of Toronto}
\affil[c]{University of Toronto Institute for Aerospace Studies}
\affil[d]{Department of Theoretical Physics and Astrophysics, Faculty of Science, Masaryk University, Czech Republic}
\affil[e]{Konkoly Observatory}
\affil[f]{California Institute of Technology}
\affil[g]{Department of Aeronautics and Astronautics, Massachusetts Institute of Technology, 77 Massachusetts Avenue, Cambridge, MA 02139, USA}
\affil[h]{Cosmic Frontier Labs}

\authorinfo{Correspondence email: coby.silayan@mail.utoronto.ca}

\begin{document} 
\maketitle

\begin{abstract}
Astronomy-grade cameras with robust performance and heritage in the space environment have long been costly, substantially limiting capacity for space-based astronomy and creating a resource barrier to access. Additionally, ultraviolet observations have historically been limited by the low quantum efficiency of most sensors in this wavelength range. The LUVCam program is designed to address both issues by providing a high-performance, low-cost, UV/optical camera system sufficiently capable to support a wide-array of space-based astronomy missions. LUVCam features a large format, low-noise, large pixel, and high quantum efficiency, commercial-off-the-shelf back(front)-side illuminated CMOS sensor, packaged with custom built readout electronics, firmware, and thermomechanical structure to provide both superlative science capability and precision on-sensor guidance at fast cadence to allow for stable high-resolution imaging. LUVCam is ITAR-free and cheap to fabricate, opening up new opportunities for access to space telescopes. Here we introduce LUVCam, describe its performance characteristics, and the rapid implementation of a technology demonstration for flight. LUVCam, coupled with a small aperture custom-built UV telescope, has been on orbit since July 2024 and has achieved Technology Readiness Level (TRL) 7. LUVCam is manifested for several more near-term orbital missions, including a second technology demonstration CubeSat for launch in 2026, and will provide both focal plane cameras for QUVIK, a two-channel UV transient astronomy mission.
\end{abstract}


\section{INTRODUCTION}
\label{sec:intro}  

Space-based astronomy is immensely consequential for our understanding of the Universe and our
place within it. Unfortunately, space-based astronomy projects have been enormously expensive – severely restricting
both who can participate and what science can be done. However, recent advances in private spaceflight have achieved greater than
an order of magnitude reduction in launch costs for typical payloads. With this launch cost reduction, a “New Space”
era has begun and brought with it dramatically cheaper and more standardized spacecrafts, space worthy components,
and satellite communication. This New Space commercial ecosystem’s quantitatively lower costs are extreme enough to represent a dramatic qualitative difference in opportunities. For instance, a half meter telescope's spacecraft bus and launch can now be procured for $<\$5$ M. However, the cost of quality astronomical instruments has not seen similar declines, which severely limits our ability to fully realize the new possibilities for access and science. The absence of a low-cost high-performance UV/optical camera system with space heritage has been identified by the community\footnote{This was a major conclusion of `Small Astronomical Space Telescopes: Exploring Cooperative Economies of Scale,' AAS Workshop, January 2023.} as one critical barrier to this dream.

The LUVCam is designed to fill this gap, as a broadly capable, multi-purpose space-grade camera to enable a critical component of low-cost astronomical instruments. Our goal is to enable a near future in which individual institutions are able to design, develop, and launch world-class telescopes at low cost and fast development timescales. This requires qualifying the latest generation of complementary metal–oxide–semiconductor (CMOS) detectors for space, selecting and qualifying components with a `careful COTS' \cite{carefulcots} approach, significantly reducing cost and time in the testing and qualification stage via automation, demonstrating functionality on orbit, and more. In this paper, we describe the first sensor selected for LUVCam and its characterization. We then describe the electronics and firmware that comprise LUVCam, before moving on to focus on the design, implementation, and launch of an orbital technology demonstration onboard the GRBBeta cubesat. LUVCam's first science demonstration at scale is to be aboard the €30M Czech national mission Quick Ultra-VIolet Kilonova surveyor (QUVIK; Werner et al.\cite{Werner22,Werner24}), but its broadly capable design is intended to enable many science cases and missions.

\section{Sensor Selection}
Many space telescopes launching today (even low-cost CubeSats) use sensors that are over 20 years old, due to their TRL and heritage. This leaves substantial performance gains on the table, as faster, more sensitive, lower noise, and higher resolution sensors rapidly become available. Often, cameras with these newer sensors are either unavailable for flight applications, or onerously expensive (see eg \cite{Gulick_2024}), significantly limiting scientific opportunities and keeping our space telescopes behind the state-of-the-art. Two things are required to change this status quo: (1) a control electronics platform that remains substantively similar (and therefore maintains its TRL) across different sensors, and (2) sufficiently low cost to enable a high flight rate to keep up with the rapidly advancing new sensor technology.

LUVCam is designed to be this general purpose sensor control platform, allowing subsequent iterations of the instrument to remain state of the art with rapidly advancing sensor technology. For the first generation LUVCam, the best sensor available on the commercial market circa 2022, that met UV capability and cost requirements, was chosen. This is the Gpixel GSENSE4040BSI(FSI), a large format back(front)-side illuminated scientific CMOS sensor, first released in 2021 with superlative broadband Quantum Efficiency (QE) and low read noise. The properties of the sensor are listed below:
\begin{table}[h!]
\resizebox{\textwidth}{!}{
\centering
\begin{tabular}{|p{4cm}|l|p{10cm}|}

\hline
 Parameter                           & Specification (high-gain vs low-gain) & Justification                                                                                                                                                                                                             \\ \hline
Peak Quantum Efficiency (QE) & 90\% @ 580 nm         & The high QE maximizes the sensitivity of the observatory. For UV imaging cases however this puts stricter requirements on the red-leak attenuation.                                                         \\ \hline 
Peak UV Quantum Efficiency (QE) (200-300nm) & 55\%          & The relatively high UV QE dramatically expands science opportunities into the UV.                                                            \\ \hline
Detector Size                 & 36.8 $\times$ 36.8mm   & \multirow{3}{10cm}{The large format and pixels of the detector enable achievable optical designs and high-resolution images.}                              \\ \cline{1-2}
Imaging Pixel Array         & 4096 $\times$ 4096    &                                                                                                                                                                                                                           \\ \cline{1-2}
Pixel size                  & 9 $\times$ 9 microns   &                                                                                                                                                                                                                           \\ \hline
Read Noise                  & 2.3 $e^-$  (high-gain)      & \multirow{2}{10cm}{The low read noise and relatively low dark current allow for both fast imaging as well as background-limited observations.}                                                                             \\ \cline{1-2}
Dark Current (-40C)           & 0.04 $e^-$/p/s    &                                                                                                                                                                                                                           \\ \hline
Full Well Capacity          & 39,000 $e^-$ (low-gain)    & \multirow{2}{10cm}{The well capacity and low read noise yields dynamic range sufficient for a variety of science goals.} \\ \cline{1-2}
Analog-to-Digital Converter                         & 12-bit  on two (high/low) gain channels      &                                                                                                                                                                                                                           \\ \hline       
Full Frame Rate & 48 Hz          & This extends capabilities to high-speed photometry. Substantially faster readout is possible for smaller Regions-of-Interest.                                                          \\ \hline
\end{tabular}
}
\caption{The technical specifications for the science sensor (GSENSE4040BSI) and the scientific justification for each relevant spec.}
\label{Tab:1}
\end{table}

\subsection{Characterization}
\label{sec:characterization}
This sensor, and its smaller sibling the Gpixel GSENSE2020BSI, were characterized in the Dunlap Institute camera characterization bench previously described by our team \cite{ajay}, and shown to meet or exceed the vendor specified quantum efficiency (QE) and noise performance.

\begin{figure} [ht]
    \subfloat{\includegraphics[width=0.5\linewidth]{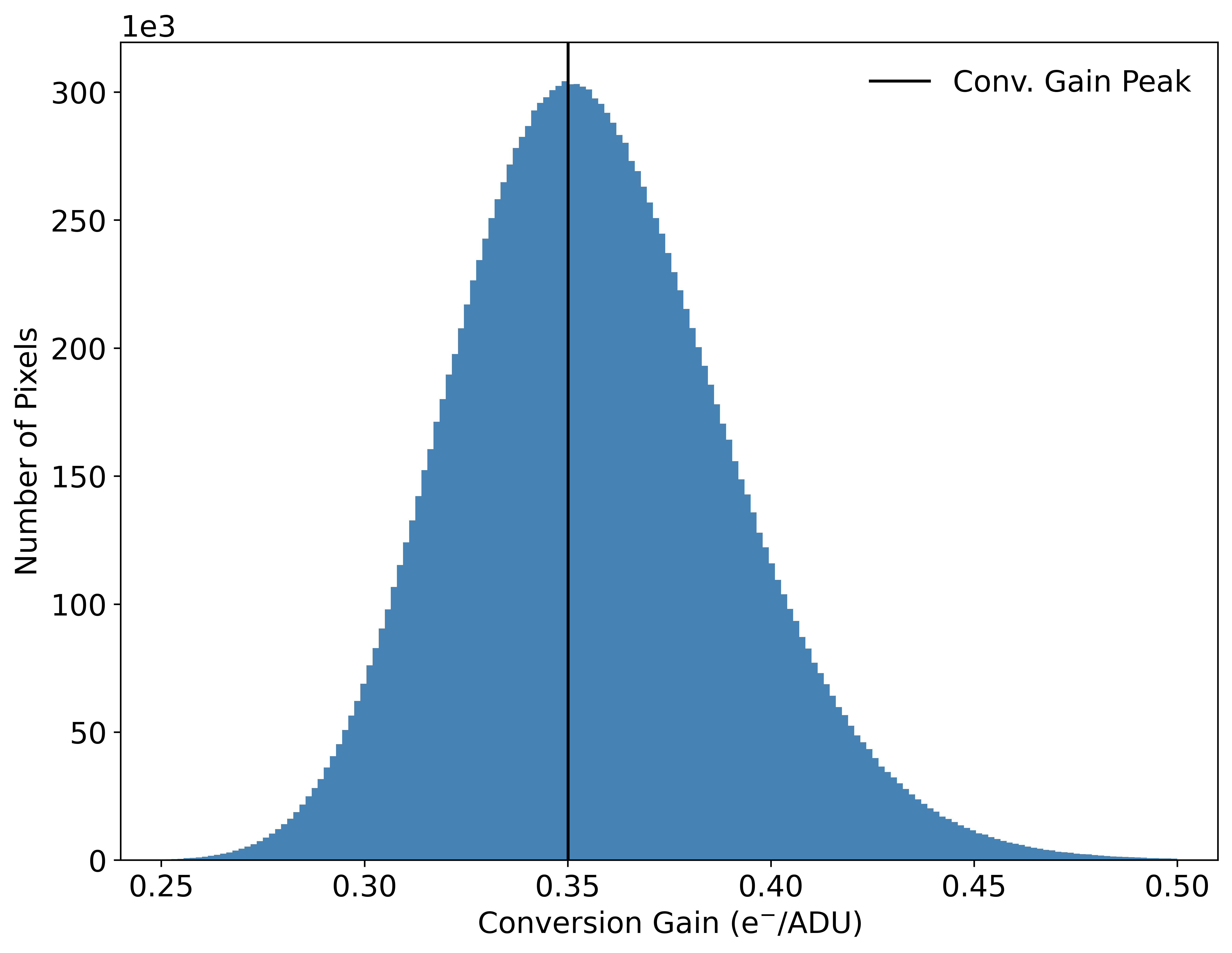}}
 \subfloat{\includegraphics[width=0.5\linewidth]{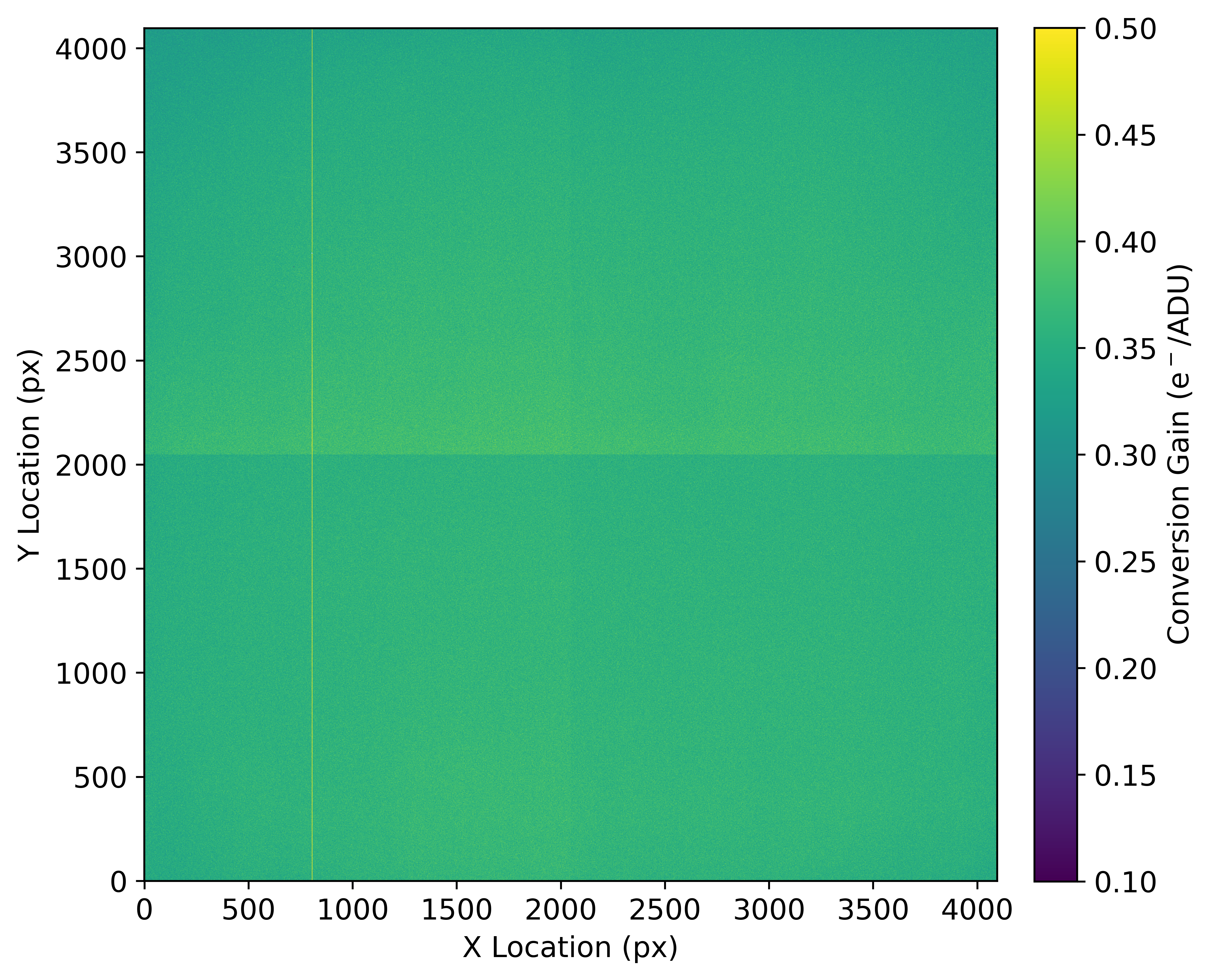}}
   \caption[The conversion gain distribution and pixel map of the GSENSE4040BSI.]{\textit{Left:} The distribution of conversion gains for each pixel in the sensor. \textit{Right:} Map of same on the sensor, note the structure which suggests a piece-wise fabrication. A defective column is also present on this sensor.}
   \label{fig:conv_gain} 
\end{figure}

\begin{figure}
    \centering
    \subfloat{\includegraphics[width=0.5\linewidth]{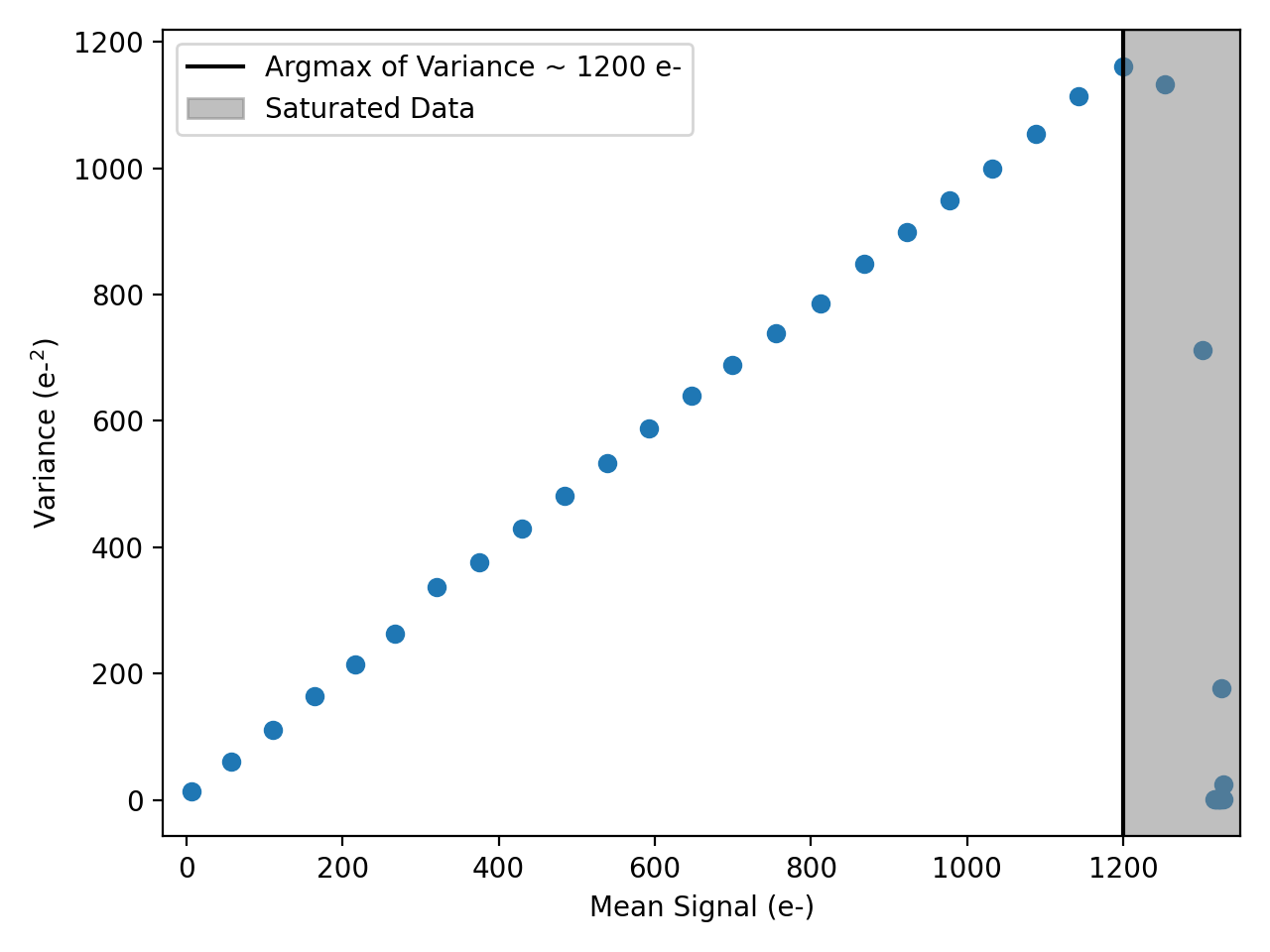}}
    \subfloat{\includegraphics[width=0.5\textwidth]{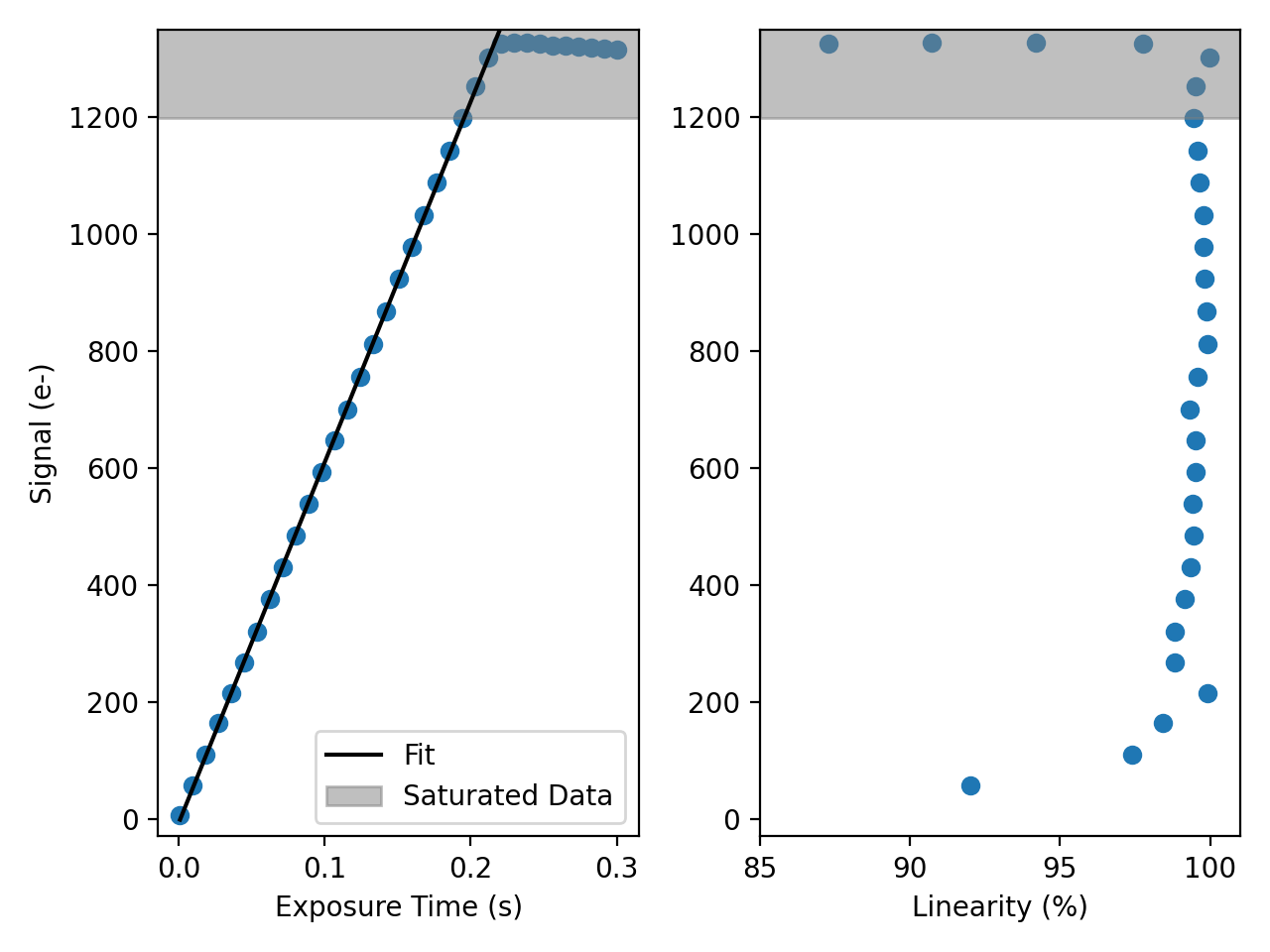}}
    \caption[The photon transfer curve and linearity of the GSENSE4040BSI.]{\textit{Left:} The photon transfer curve for the GSENSE4040BSI sensor, showing saturation around $\sim1220$ $e^-$ in high gain mode. \textit{Right:} The signal vs. exposure time and linearity. The linearity is $>98\%$ in the shot-noise dominated regime.}
    \label{fig:PTC-linearity}
\end{figure}

The image sensors support simultaneous readout of both low and high gain images, with different conversion gain, read noise, and full-well characteristics. Characterization focused on the high gain mode. To measure the per-pixel conversion gain of the sensors, they were exposed to diffused light, and both light and dark images were taken at exposure times yielding a median signal ranging from the bias level up to the saturation limit of the detector. For each exposure time/signal level, a stacked dark image was created. The light images were normalized to the first image of the set to account for variations in illumination. The stacked dark image was then subtracted, and these dark-subtracted images were mean and variance-stacked to create a pixelwise mean-variance dataset. The conversion gain was obtained for each pixel's data following \cite{janesick}; for the GSENSE4040BSI, the typical conversion gain in high gain mode was $0.35\pm0.05$ $e^-$/ADU, see Fig. \ref{fig:conv_gain}. Photon transfer curves were also constructed, and the saturation of the sensor in high-gain mode was found to be $\sim1200$ $e^-$, with linearity $>98\%$ in the shot-noise dominated region as shown in Fig. \ref{fig:PTC-linearity}.

\begin{figure} [ht]
   \subfloat{\includegraphics[width=0.5\linewidth]{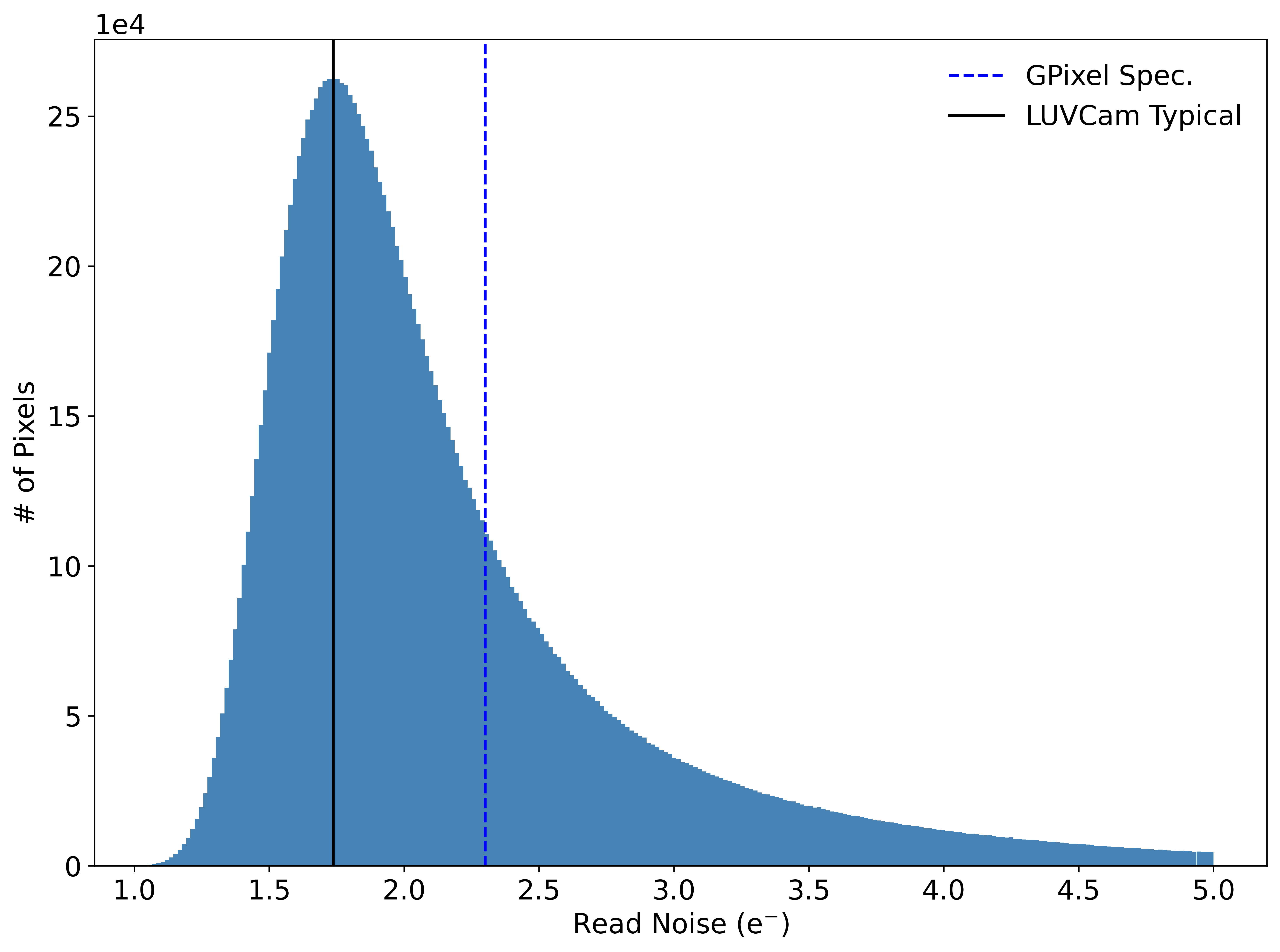}}
   \subfloat{\includegraphics[width=0.5\linewidth]{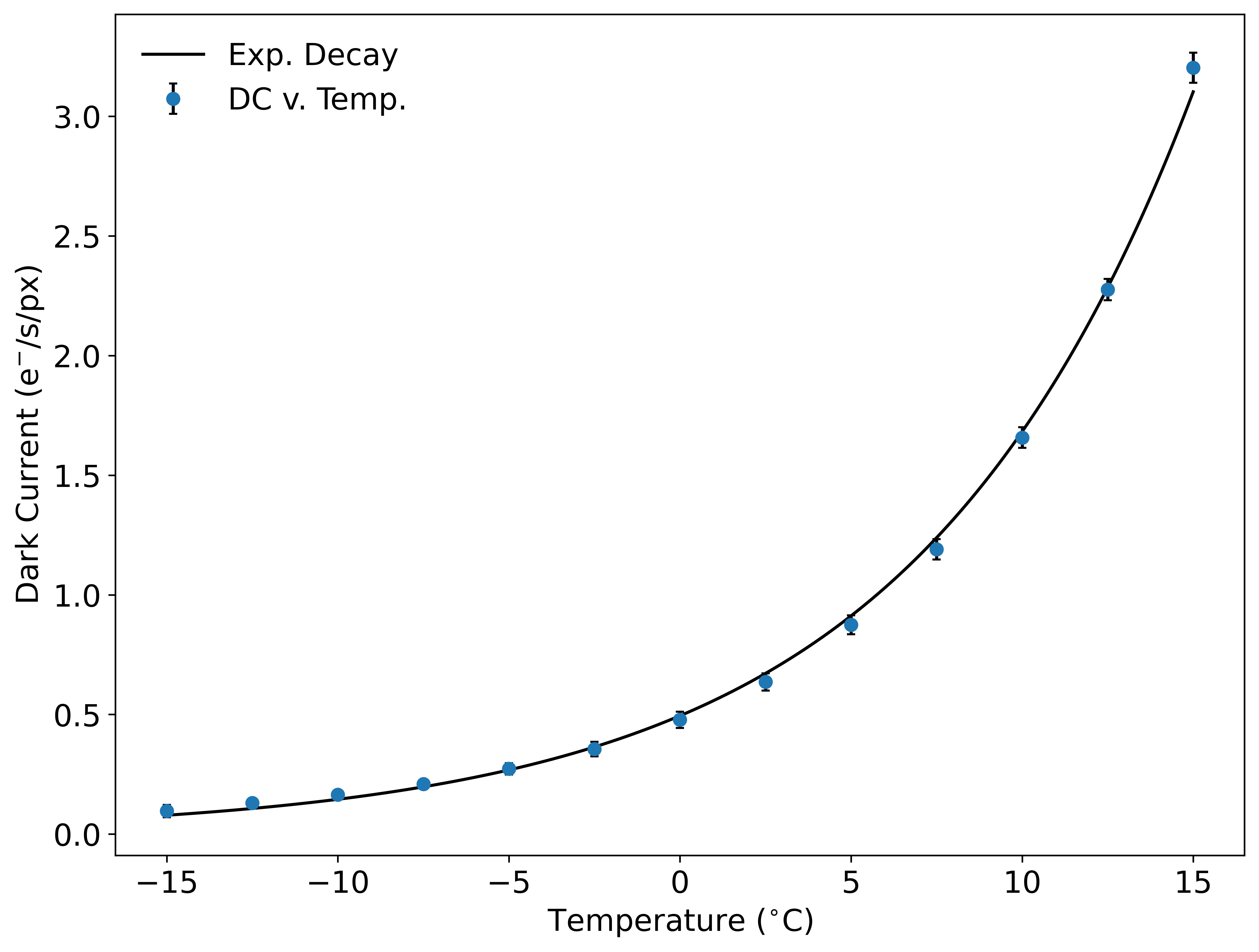}}
   \caption[The read noise distribution, and dark current as a function of temperature, of the GSENSE4040BSI]{\textit{Left:} The read noise distribution of the Gpixel GSENSE4040BSI. These data have been multiplied by the conversion gain map to convert the read noise from ADUs to electrons. \textit{Right:} The dark current as a function of sensor die temperature showing a clear exponential decay to at least -15 C. }
   \label{fig:RNDC} 
\end{figure}

The read noise was calculated by standard-deviation stacking 1000 bias frames, then measuring the peak of the distribution of values across all pixels. The GSENSE4040BSI sensor had a typical (mode) read noise of 1.7 $e^-$, overperforming the Gpixel spec which is closer to the mean value of 2.3 $e^-$. These results are presented in Fig. \ref{fig:RNDC}, left.

To characterize the dark current, 10 dark exposures and biases were taken at exposure times ranging from 0.1 s to 100 s, at temperatures from -15 \textdegree C to 15 \textdegree C in 2.5 \textdegree C steps.
The mean value of the bias-subtracted stacked frame for each exposure time is calculated. These mean values are then plotted as a function of the corresponding exposure time, and the slope of a line fit to these data is taken to be the dark current. Since the mean values of the full frame are used to obtain the dark current, the typical conversion gain value is used to report the dark current in units of $e^-$/s. The average dark current of the GSENSE4040BSI sensor at -15 \textdegree C was $0.10\pm0.03$ $e^-$/s/px. As expected, the measured dark current exponentially decayed with decreasing sensor temperature, and no plateauing was seen down to -15$^\circ$C, shown in Fig. \ref{fig:RNDC} right. We intend to extend this dark current characterization to colder temperatures in the near future, as apparent DC plateaus due to eg proximity electronics hot-spots producing trapped IR photons are known to exist in similar sCMOS sensors.

The characterization equipment used to measure the QE included a stabilized Xenon lamp, monochromator, integrating sphere, and calibrated photodiode \cite{ajay}. The dominant source of statistical uncertainty in the QE originates in the conversion gain measurement. We note that the systematic uncertainty in absolute QE normalization arising from this characterization bench is not yet fully understood, and future work remains to stabilize the system and make repeatable measurements. This QE measurement is not corrected for the increased quantum yield of photons below $\sim300$ nm. Some previous works\cite{quantumyield} find this is a $\sim30\%$ relative effect by 200 nm, although the results are often sensitive to the specific properties of the silicon. The quantum yield for our camera will be directly measured in the near future.

\begin{figure} [ht]
    \centering
   \includegraphics[height=8cm]{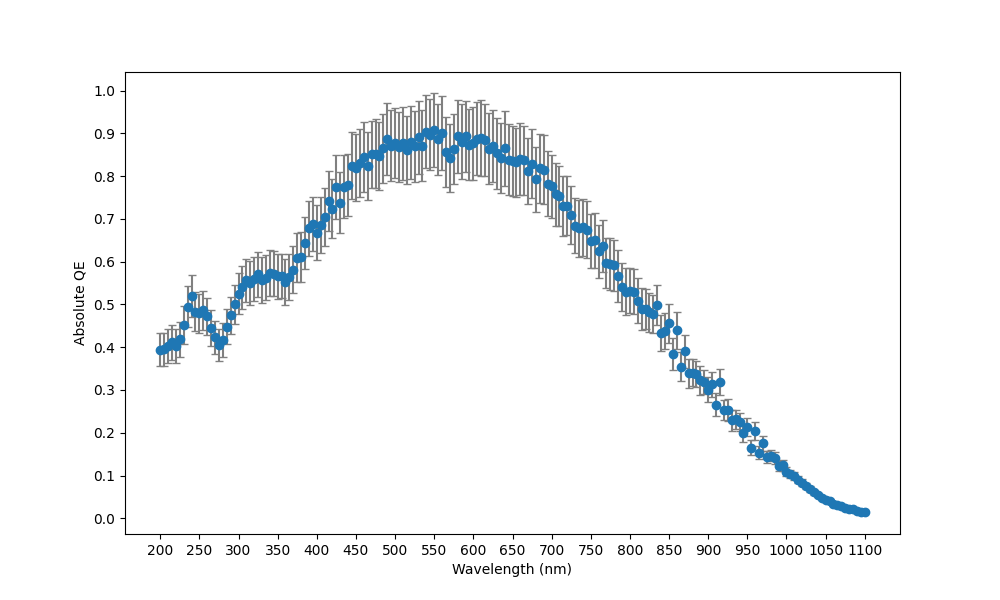}
   \caption[The absolute quantum efficiency of the GSENSE4040BSI]{The absolute quantum efficiency of the GSENSE4040BSI. This curve is not corrected for enhanced quantum yield at wavelengths below 300 nm.}
   \label{fig:qe}
\end{figure} 

\subsubsection{Effect of radiation}
To better guide a custom camera design, and obtain a baseline understanding of the expected impact of radiation on the sensor performance, we performed accelerated radiation testing at the TRIUMF Proton Irradiation Facility's 105 MeV proton beam-line. For this test the GSENSE2020BSI (a smaller version of the 4040 with similar sensor architecture, pixel structure, epitaxial layer and AR coating) was tested in a COTS camera whose pre-radiation characterization we presented in \cite{ajay}.

\begin{figure}[h!]
\centering
\begin{minipage}{.4\textwidth}
    \begin{tabular}{lcc}
    Min proton energy & 63       & MeV               \\
    Max proton energy & 105      & MeV               \\ 
    \multicolumn{3}{c}{\textbf{Standard Beam}}               \\ \hline
    Min flux          & $10^5$ & proton/$\mathrm{cm}^2$/s \\
    Max flux          & $10^8$ & proton/$\mathrm{cm}^2$/s \\
    Dose Rate (Si)    & 7.5      & rad/s           \\
    Spot Size  & $5\times 5$& cm \\
    \multicolumn{3}{c}{\textbf{Upstream Beam}}                \\ \hline
    Max flux          & $2\times 10^9$ & proton/$\mathrm{cm}^2$/s \\
    Dose Rate (Si)    & 100      & rad/s     \\
    Spot Size & 2 & cm diameter circle 
    \end{tabular}
    \vspace{1.5cm}
    \captionof{table}{A summary of the relevant TRIUMF beam specifications for both the standard and upstream configurations.}
    \label{tab:triumf_summary}
\end{minipage}%
\hspace{1.5cm}
\begin{minipage}{.5\textwidth}
  \centering
    \includegraphics[width=\linewidth]{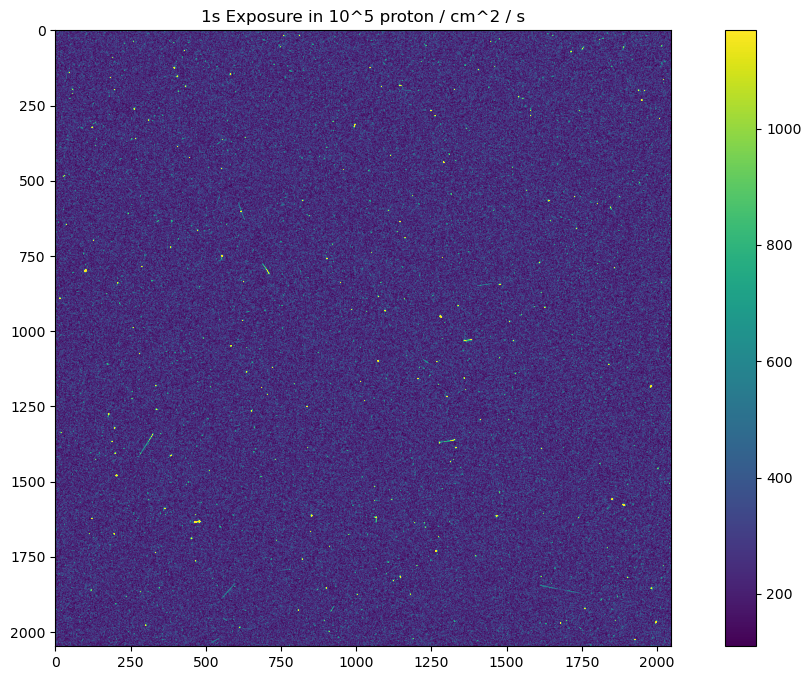}
    \caption{A 1 second exposure during beamline testing at minimum beam flux (SAA passage equivalent).}
    \label{fig:imginrad}
\end{minipage}
\end{figure}

Before the start of radiation testing, five images were taken using the low-gain readout mode at exposures times of 0.1, 1, 10, 30, 60, and 300 seconds, with the sensor at 17\degree\ C. The camera was then placed in the BL2C beam-line at TRIUMF, see Tab. \ref{tab:triumf_summary}. The testing was conducted in three parts: 1) a radiation dose of 1650 rad over approximately 10 minutes with minimum beam flux (Standard), which coincidentally corresponds to the maximum flux experienced during a 550 km altitude passage through the South Atlantic Anomaly (SAA) 2) an additional dose of 1650 rad at higher beam flux (Standard) over approximately 3 minutes and 3) a cumulative dose of 100 krad at maximum beam flux (Upstream) in under one hour. 1650 rad TID was chosen as a baseline target to simulate a 5 year LEO mission with 5 mm of aluminum equivalent shielding. 100 krad TID is equivalent to 5 years in a sun-synchronous LEO with no shielding, or an Io exploration mission in an inclined Jupiter orbit with 2.5 mm of Al-equivalent shielding \cite{jupitercam}.

\begin{figure} [h!]
       \subfloat{\includegraphics[width=0.5\linewidth]{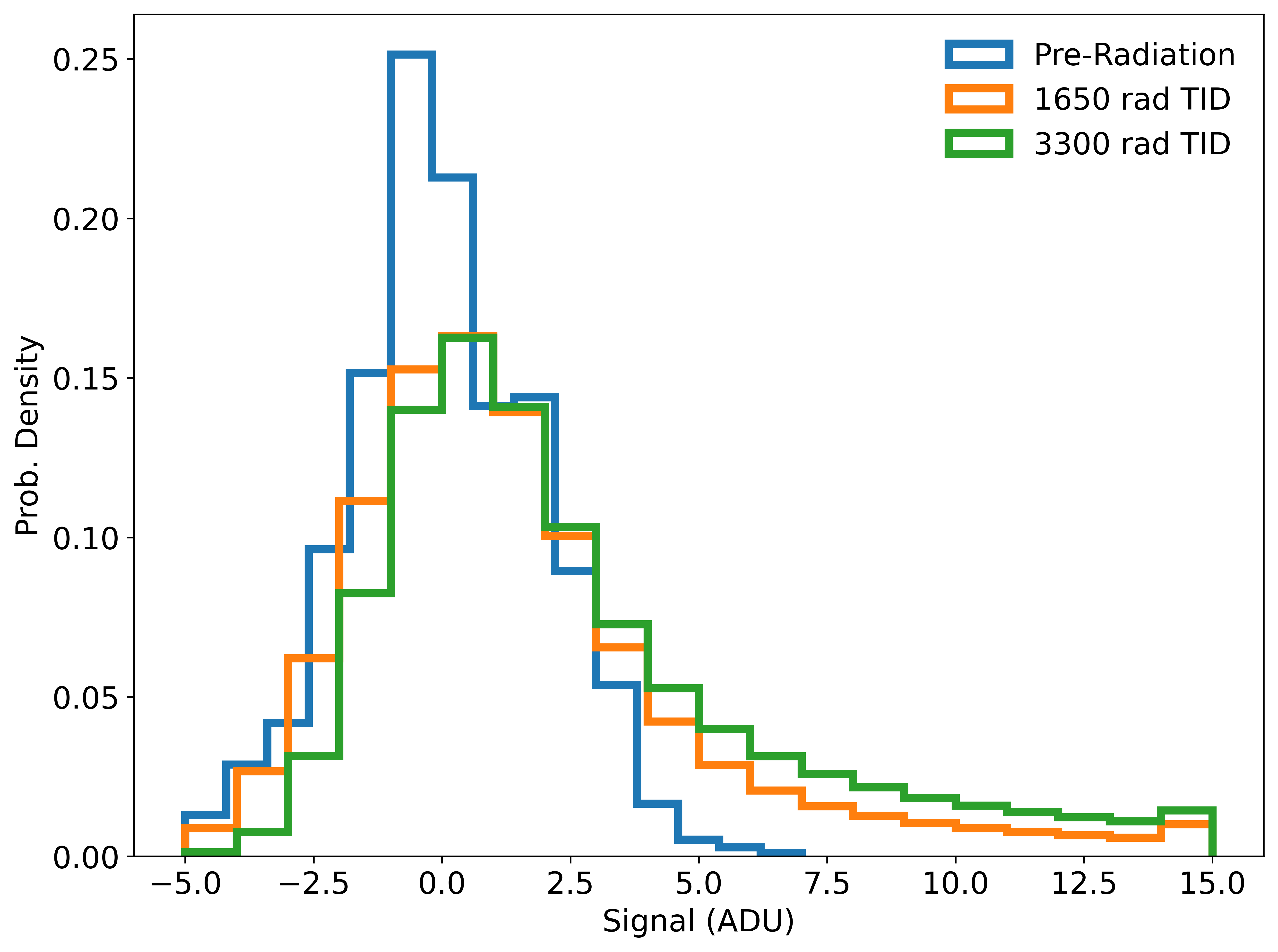}}
   \subfloat{\includegraphics[width=0.5\linewidth]{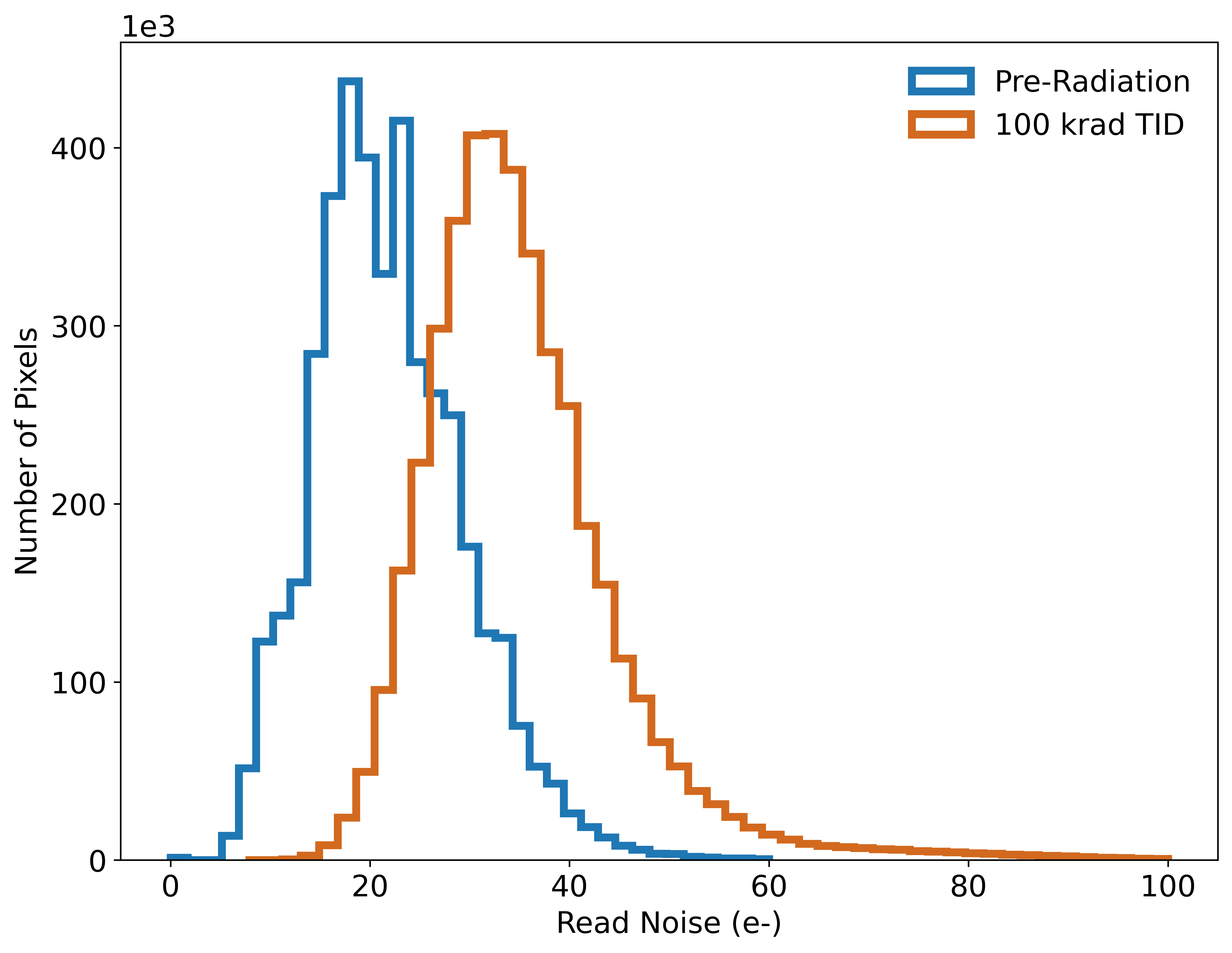}}
   \caption[Read-noise distribution of the GSENSE2020BSI before and after radiation testing.]{The read noise distribution of the GSENSE2020BSI in low-gain mode before and after radiation.}
   \label{fig:postradRN}
\end{figure}

The lack of sufficient data taken between the 1650/3300 rad and 100 krad runs prohibits direct measurement of the read-noise degradation after 1650 and 3300 rad. Here we make a relative comparison of the distribution of the mean/mode subtracted data as proxy for the read noise distribution. After the sensor experienced an initial TID of 1650 rad, a noticeable degradation in read noise occurred, with the distribution shifting away from the peak toward higher values, see left panel of Fig. \ref{fig:postradRN}. A significant high read noise tail emerges, expanding the distribution range from approximately -5 to 7 ADU in the pre-radiation data to -5 to 15 ADU. Since these are 0.1-second dark exposures with the mode of the distribution subtracted, changes in the bias level are accounted for, and the dark current is expected to be negligible. This indicates that the shift in the distribution is indeed due to read noise degradation. The trend continues in the 3300 rad TID data, with more pixels showing higher read noise, though the difference between the 3300 rad and 1650 rad TID distributions is less pronounced than between the 1650 rad and pre-radiation distributions. 
Access to the camera again after 100 krad TID allowed direct measurement of the read noise, following the pre-rad method. We measure an approximate doubling of the sensor read noise after 100 krad TID.

\begin{figure}
    \centering
    \includegraphics[width=0.5\linewidth]{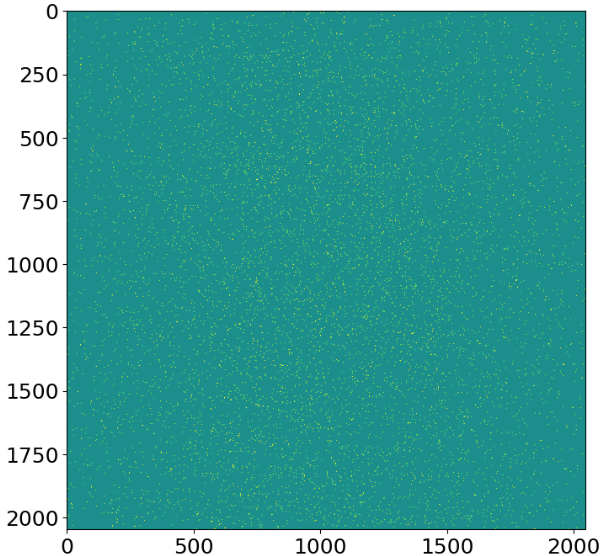}
    \caption{Binary map of pixels flagged as non-linear after 100 krad TID.}
    \label{fig:post-rad-linearity}
\end{figure}

The linearity was also measured post 100 krad TID, and we generally find little to no substantial damage to the portion of the well relevant to linearity. In Fig. \ref{fig:post-rad-linearity} we show the pixels that exhibit behavior $>3\times$ the mean standard error of a linear fit. We flag these pixels as non-linear. They make up $<0.2\%$ of the entire array and are randomly distributed across the sensor.

The post-100 krad TID dark current data yields a mean dark current of 40.26 $\pm$ 0.05  $e^-$/s/pix at -30 $^{\circ}$C, a factor of $\sim$60 larger than the 0.7  $e^-$/s/pix measured pre-radiation. This finding is consistent with others in the literature that find comparable or even larger differences after sensor irradiation, though it should be noted that different types of energetic particles can affect this result \cite{darkcurrent_100x}. Given that dark current scales linearly with TID~\cite{dc_linear_with_dose}, the expected dark current increase after 1650 rad TID is $< 2\times$, corresponding to a 0.13 $e^{-}$/s/pix per year increase behind 5mm of aluminum. Given the dark current profile with temperature in Fig. \ref{fig:RNDC}, this implies an extra 6\degree\ of cooling required by end of 5 year mission to maintain DC at mission start levels. Alternatively, $3.95\times10^{-4}$  $e^-$/pix/s increased dark current per rad, for shielding design.

 \begin{figure}
     \centering
     \subfloat{\includegraphics[width=0.5\linewidth]{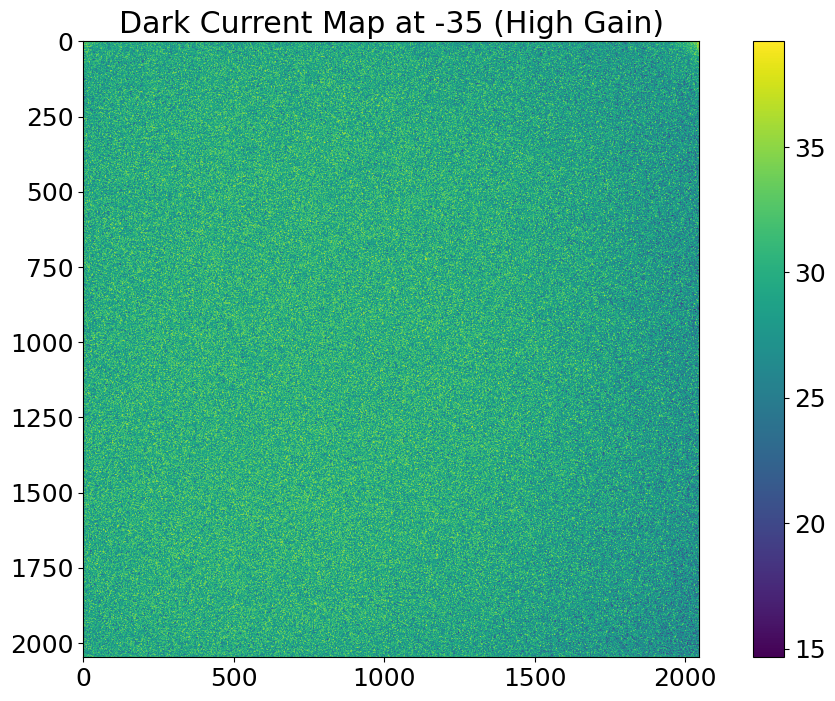}}
     \subfloat{\includegraphics[width=0.5\linewidth]{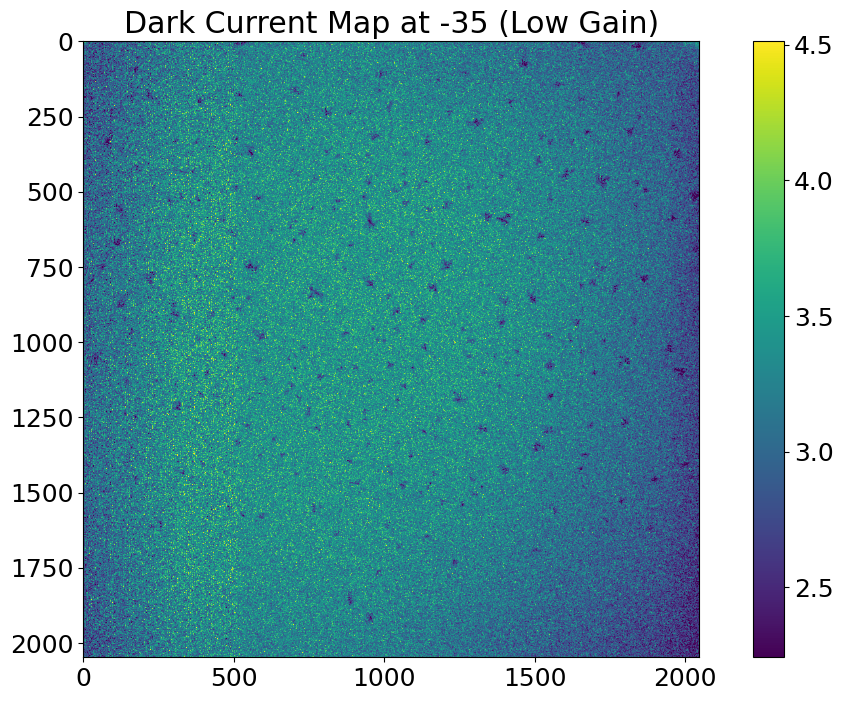}}
     \caption[Post-irradiation dark current map at -35 C.]{The dark current map in ADU/second as a function of pixel location at -35C after 100 krad TID. Note that some of the post radiation features are only visible in the low gain channel.}
     \label{fig:postradDCmap}
 \end{figure}
 
Overall, the effects of radiation on $<10$ year LEO mission timescales are manageable, indicating that these sensors will remain operable and are suitable for space applications. No beam test is perfect however, and it is known that dose rate is important to radiation induced degradation, not just TID. 

These points justify in-orbit demonstration. 

\section{Implementation}
\subsection{Electronics}
The LUVCam is implemented in two parts, a control module and sensor module, both Printed Circuit Boards (PCBs) in PC104 format to allow easy integration in a CubeSat form-factor, and connected via flex cables. The flex connection allows the sensor and camera electronics to be separated for appropriate shielding and thermal control. Overall power management for the system is provided by a power management integrated circuit (PMIC) in the electronics module with independent control pathway from the on-board computer. This ensures the camera control system can be restarted should there be an in-flight soft error in the control electronics. Operating the CMOS sensor requires stable power management, high-bandwidth data transfer, impedance matched traces, and high-fidelity clocks. The GSENSE sensor has a 140-pin Pin Grid Array (PGA) electrical interface that the sensor module must appropriately seat and route power, biases, control signals, and 18 pairs of low-voltage differential signaling (LVDS) channels each running at 600 MHz. A Field Programmable Gate Array (FPGA) was chosen to control the camera to allow for a re-programmable platform and the Xilinx Artix-7 XC7A100T-2CSG324I was selected due to its large number of logical elements and I/Os which allow ample room for the platform to support future, larger, sensors. DDR memory for buffering images and a configuration flash for storing FPGA firmware are included in the control module. 

A high level block diagram of the system is shown in Fig. \ref{fig:diagram}.

\begin{figure}[ht]
    \centering
    \includegraphics[width=0.75\textwidth]{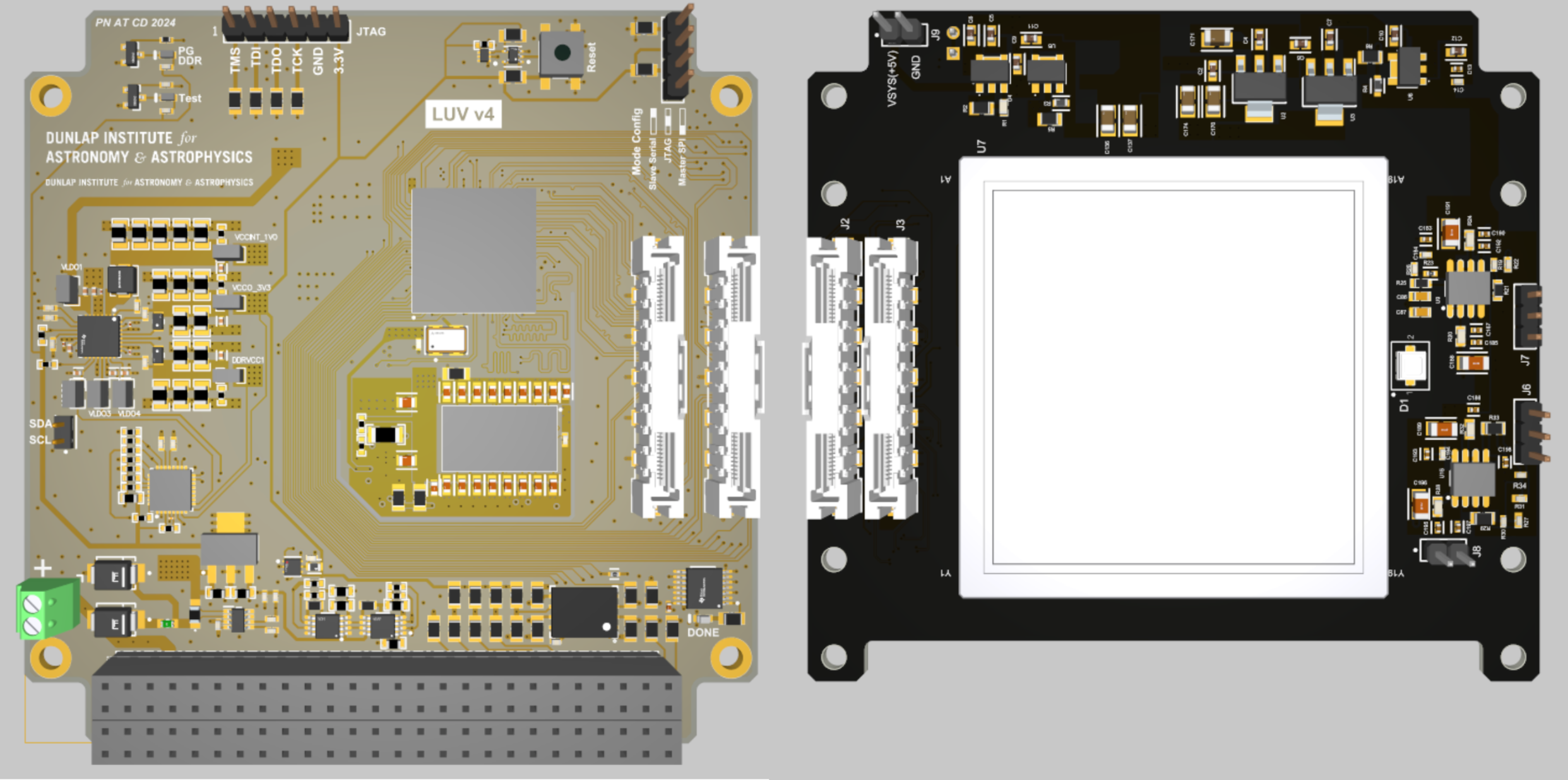}
    \caption[A render of the LUVCam electronics.]{Top-side of the control electronics module (left) and sensor module (right) in their PC104 implementation for the CubeSat Tech Demo. Flight version is without pin terminals. Flex cable is not shown.}
    \label{fig:boardmodel}
\end{figure}

Commercial and industrial grade electronic components were selected with sensitivity to cost and availability. Careful selection was made to avoid electrolytic capacitors (replaced with ceramic) and components with high tin content due to their possible failure modes under vacuum. No `space-grade' or `rad-hard' components are included in the design, and instead a `careful COTS'\cite{carefulcots} approach is taken to development.

\subsection{Firmware} 
\begin{figure}[ht]
    \centering
    \subfloat{\includegraphics[width=0.5\textwidth]{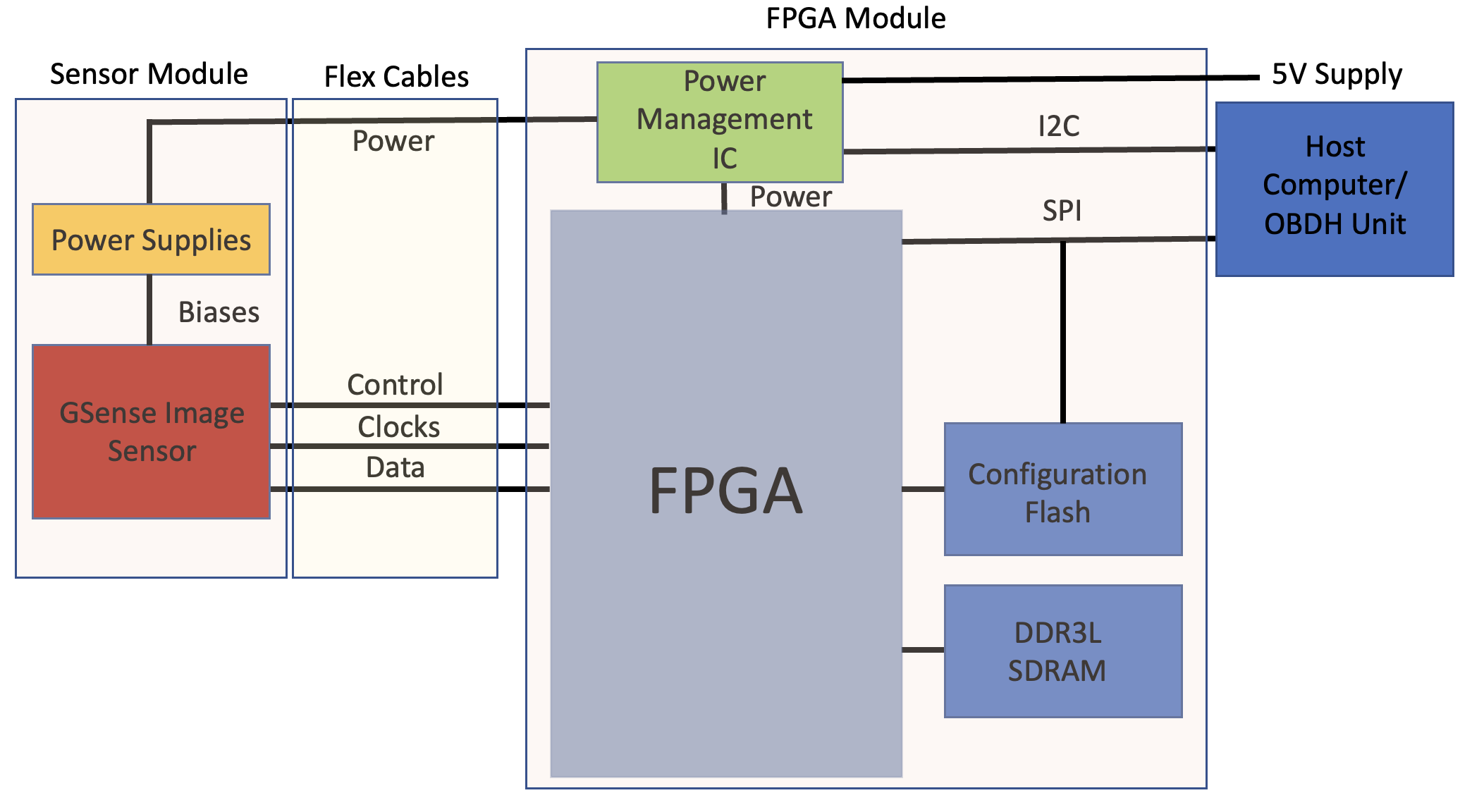}}
    \subfloat{    \includegraphics[width=0.5\textwidth]{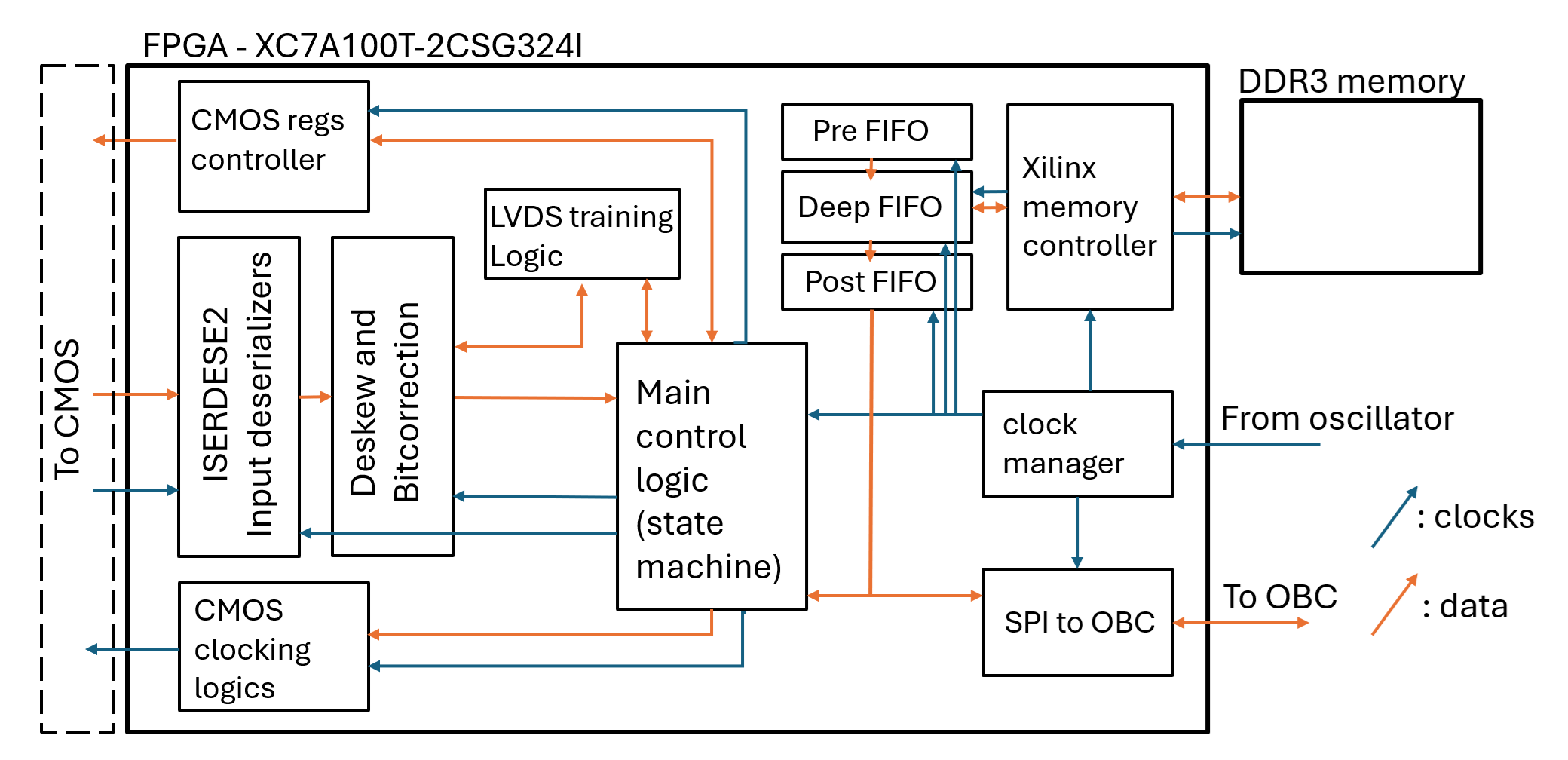}}
    \caption[Block diagram of the LUVCam system and FPGA firmware.]{\textit{Left:} Block diagram of camera electronics. \textit{Right:} Simplified FPGA Firmware logic diagram of LUVCam.}
    \label{fig:diagram}
\end{figure}

LUVCam's FPGA is responsible for controlling the CMOS sensor, capturing raw data, buffering it, and transmitting the captured frames to the On-Board Computer (OBC). A simplified logic diagram of the firmware is shown in Fig. \ref{fig:diagram}, right.

\paragraph{CMOS Sensor Interfaces} The CMOS sensor communicates with the FPGA via three distinct interfaces: \begin{itemize} \item A modified Serial Peripheral Interface (SPI) interface for configuration register read/write operations. \item Timing signals for internal pixel control. \item 16-channel LVDS interface for data readout. \end{itemize}

The SPI interface is managed by a simple state machine, which takes the desired values from the OBC commands, reformats them into the appropriate SPI bitstream, and updates the sensor registers accordingly. The registers control various settings within the sensor. The timing signals consist of predefined bit sequences provided by the CMOS manufacturer, which we hardcoded into a Look-Up Table (LUT). These signals are written to the camera in the correct sequence during data acquisition and readout. In addition to the timing signals, a 50 MHz clock signal is supplied to the sensor as its primary clock. The digitized output data from the sensor is transmitted as serial bitstreams via 16 data channels over the LVDS interface, along with a pair of differential clock signals (DCLK) to the FPGA.

To account for LVDS trace length differences in the PCB, we implemented a training algorithm that artificially adds delays on shorter lines. The sensor first outputs a fixed pattern of bits, and the FPGA dynamically adjusts the delay on each channel to ensure data sampling aligns with the rising and falling edges of the DCLK. The ISERDESE2 IP is used to convert the serial bitstreams into 6-bit words, followed by 2:1 deserialization into 12-bit words. A word-alignment algorithm based on the training word is integrated into the gearbox to correct potential bit-shifts \cite{AMD_UG953, lau2022development}.

\paragraph{Data Buffering} Since the connection between the FPGA and OBC can be slow, and the OBC may have limited memory, captured image data from the sensor must be temporarily stored in DDR3 memory (2 Gbit) connected to the FPGA, managed by Xilinx's memory controller IP. This is sufficient to hold 10 images from the 4096 x 4096 pixel, 12-bit, sensor.

We implemented a multi-stage FIFO buffer using DDR3 memory on the FPGA board and DeepFIFO IP \cite{xillybus2019deepfifo}. As the DDR3 memory is external, it experiences read/write delays of several clock cycles. To prevent back-pressure and data loss, a small pre-FIFO buffer is used to temporarily store incoming data when the memory is busy. Once the memory becomes available and the pre-FIFO is not empty, data is transferred from the pre-FIFO to the DDR3 memory through the DeepFIFO logic. A post-FIFO buffer is also used to smooth the data output. When the post-FIFO is half-empty, the DeepFIFO logic triggers the release of data from the memory, storing it in the post-FIFO for access by downstream logic.

\paragraph{Communication Interface} Communication and control of the camera FPGA is implemented in a single SPI interface, where the OBC controls the FPGA. Like the modified CMOS-FPGA SPI interface, the communication is implemented using a state machine architecture. A simple command set is programmed on the FPGA, enabling the OBC to send commands to control the FPGA, take an exposure, and receive image data in return. All control commands from the OBC are 32 bits long (8 hex digits), while the data received from the FPGA can vary in length, up to the contents of the DDR3 memory. 




\section{Technology Demonstration}
A low-cost, rapid, on-orbit test and flight demonstration was determined to be most appropriate for the design and development philosophy of LUVCam. This approach, particularly when followed by further iterative flight testing, can yield the fastest and highest performance final product, while avoiding long, slow, and expensive testing and qualification campaigns on the ground. In particular, the robustness of the system to the launch and space environment, including vibration, g-loading, shock, ultra-high vacuum, thermal cycling, and radiation can be immediately established. In addition, the sensor performance and degradation as a function of orbital lifetime can be measured, and used to baseline its use for future larger missions. However, the `best test is a flight test' approach obviously requires getting to space cheaply.

The GRBBeta spacecraft is a 2U CubeSat initially designed for detection of gamma-ray bursts, and to help develop technology for future gamma-ray detecting constellations. In November 2022, GRBBeta was awarded a launch slot on the maiden flight of the Ariane 6 rocket, with 2U of volume available in the deployer. The GRBBeta payload is very similar to that on GRBAlpha \cite{GRBAlpha}, comprising a CsI(Tl) gamma-ray scintillator coupled to a flat Silicion Photo-Multiplier (SiPM) for readout, yielding sensitivity to photons from 70-890 keV, with a peak effective area of $\sim55$ cm$^2$. GRBAlpha was a 1U CubeSat; the extension of GRBBeta to 2U opened up volume and mass for another payload, and allowed an opportunity for a low-cost, fast-turnaround LUVCam technology demonstration on orbit. 

\begin{figure}
    \centering
    \subfloat{\includegraphics[width=0.5\linewidth]{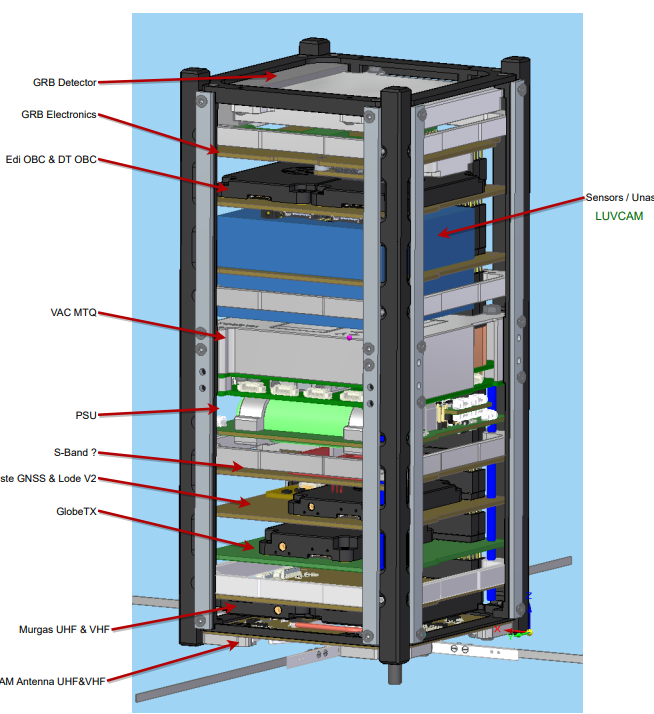}}
    \subfloat{\includegraphics[width=0.472\linewidth]{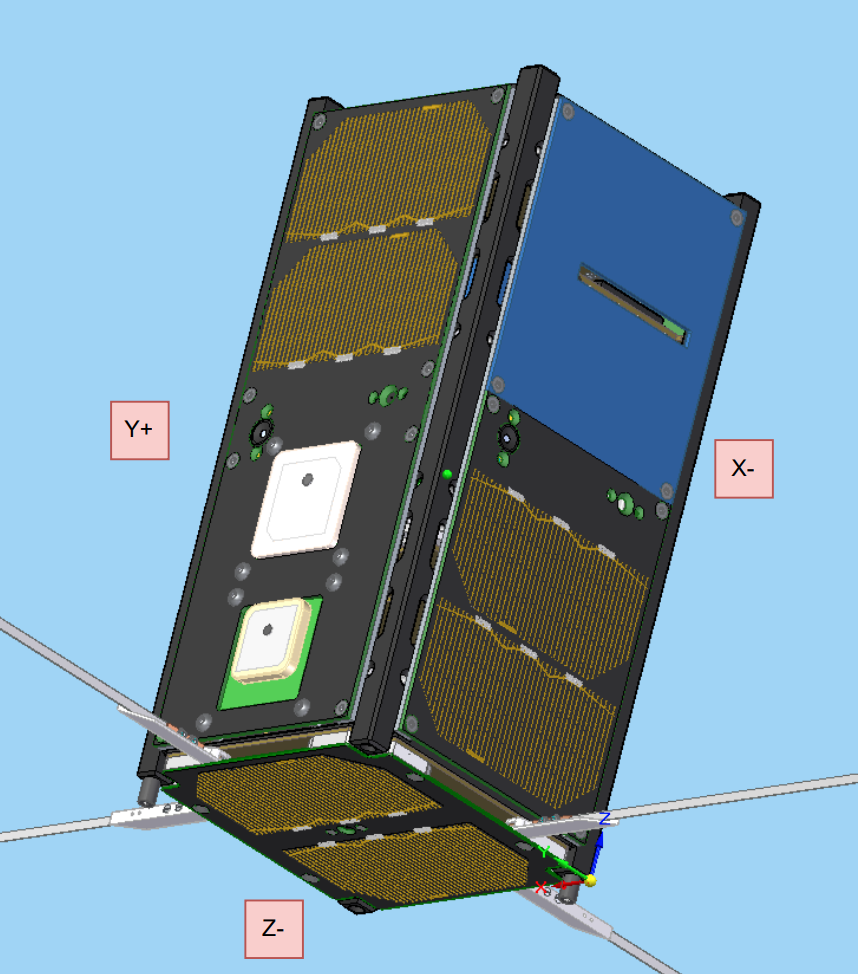}}
    \caption[The free-volume available in the GRBBeta spacecraft for LUVCam.]{\textit{Left}: The free volume available in the GRBBeta Cubesat is shown in blue, $\sim$ 96 x 96 x 40 mm$^3$. \textit{Right:} One side-panel of the GRBBeta spacecraft (-X,+Z) was free of photovoltaic cells or antennae, and available for use as long as it provided an external access slot to the OBC and power supply interface.}
    \label{fig:freevolume}
\end{figure}

The volume available in early 2023 for a secondary payload is shown in Fig. \ref{fig:freevolume}, and comprises $\sim$ 96 x 96 x 40 mm$^3$ in the upper (+Z) U of the CubeSat, sitting directly below the On-board Computer (OBC) and above the Attitude Determination and Control System (ADCS aka VAC). Less than 300 grams of mass allocation was available for a secondary science payload.

The small mass and volume allotment, along with the rapid launch schedule ($<$ 1 year from design start to delivery), modest budget, and small team, were challenging constraints on the  design, fabrication, and testing process. To ensure success, relatively modest goals were set for this first flight. The defined goal of the mission was to: \textit{Demonstrate operation of LUVCam on orbit, verifying noise performance (read, dark), radiation robustness, and launch + thermal cycling survival at a data quality and price point that allows qualitatively new opportunities for future missions}. This was translated into a set of Level 0 Mission requirements for the LUVCam Tech Demo flight on GRBBeta in Tab. \ref{tab:L0}.

\begin{table}[ht]
\resizebox{\textwidth}{!}{%
\begin{tabular}{llll}
{\color[HTML]{1F497D} \textbf{ID}} & {\color[HTML]{1F497D} \textbf{Name}}            & {\color[HTML]{1F497D} \textbf{Requirement Text}}                                                                                                                                                           & {\color[HTML]{1F497D} \textbf{Rationale}}                                                                                                                                       \\ \hline
\multicolumn{1}{|l|}{L0-REQ-010}   & \multicolumn{1}{l|}{First, do No Harm}          & \multicolumn{1}{l|}{\begin{tabular}[c]{@{}l@{}}LUVCam failure shall have no failure modes that cascade into failures of \\ the GRBBeta mission.\end{tabular}}                                              & \multicolumn{1}{l|}{Our design ethic.}                                                                                                                                          \\ \hline
\multicolumn{1}{|l|}{L0-REQ-020}   & \multicolumn{1}{l|}{Data Recording and Storage} & \multicolumn{1}{l|}{\begin{tabular}[c]{@{}l@{}}LUVCam shall allow sensor data and temperature telemetry to be recorded, \\ stored, and transmitted to the OBC on command.\end{tabular}}                    & \multicolumn{1}{l|}{The primary goal of the mission is to operate the camera in orbit.}                                                                                         \\ \hline
\multicolumn{1}{|l|}{L0-REQ-030}   & \multicolumn{1}{l|}{Characterization}           & \multicolumn{1}{l|}{LUVCam shall characterize the sensor on orbit.}                                                                                                                                        & \multicolumn{1}{l|}{\begin{tabular}[c]{@{}l@{}}Noise performance (read, dark), image quality (flat), radiation hardness,\\  and launch + thermal cycling survival.\end{tabular}} \\ \hline
\multicolumn{1}{|l|}{L0-REQ-040}   & \multicolumn{1}{l|}{GRBBeta Payload}            & \multicolumn{1}{l|}{\begin{tabular}[c]{@{}l@{}}LUVCam shall mount to the GRBBeta structure, fit within payload \\ mass/volume/power allocation, and comply to the Interface Control Document.\end{tabular}} & \multicolumn{1}{l|}{System must fit in alloted space or no launch.}                                                                                                             \\ \hline
\multicolumn{1}{|l|}{L0-REQ-050}   & \multicolumn{1}{l|}{Survival and Operations}    & \multicolumn{1}{l|}{LUVCam shall survive launch and on-orbit operational conditions.}                                                                                                                       & \multicolumn{1}{l|}{System must survive launch and work in orbit.}                                                                                                               \\ \hline
\multicolumn{1}{|l|}{L0-REQ-060}   & \multicolumn{1}{l|}{Cost}                       & \multicolumn{1}{l|}{The system shall not cost more than \$50 K CAD, not including NRE.}                                                                                                                    & \multicolumn{1}{l|}{Our budget based on sponsor funding, and goal of low-cost replicability.}                                                                                   \\ \hline
\multicolumn{1}{|l|}{L0-REQ-070}   & \multicolumn{1}{l|}{Life}                       & \multicolumn{1}{l|}{LUVCam shall operate for at least 90 days on orbit}                                                                                                                                    & \multicolumn{1}{l|}{To allow for characterization con-ops and radiation dose.}                                                                                                  \\ \hline
\end{tabular}%
}
\caption{The L0 mission requirements, or Sacred Expectations, for the LUVCam Tech Demo flight.}
\label{tab:L0}
\end{table}

\subsection{Let there be light}
The LUVCam Tech Demo mission requirements are modest, and can in principle be met with the two LUVCam PCBs (controller and sensor) stacked vertically supported on the CubeSat rails (as the other payloads are in Fig. \ref{fig:freevolume}), connected to each other via flex cable, and with the sensor contained in a small light-tight enclosure with a supplementary LED to provide diffuse flat-field illumination. This approach is very simple, requires minimal mechanical engineering, and focuses the effort on the fundamental electrical subsystems, the heart of the camera. Indeed, this was the original plan. However, during early development it was noted that of the 10 side-panels comprising the GRBBeta exterior, one was free of photovoltaic cells or antennae, due to its use providing an access slot for post-integration communications, software checks, and battery charging (see Fig. \ref{fig:freevolume}, right). This panel was adjacent to the free volume, and could be replaced, as long as it still met the requirements of providing an access slot, and integrating with the deployer.

With the opportunity for a path to free space, this group of astronomers could not resist attempting to image the UV sky. However, this \textit{significantly} complicates the mission design. Imaging requires optics (most of the time, if your electronics are slow). As the sensor itself is too large to orient along the Z-axis of the spacecraft in the given free volume, this meant designing, fabricating, testing, and integrating a custom, folded, UV telescope. The downstream requirements of imaging and focal length stability drove the need for an independent mechanical frame, rigidly supporting the optics and sensor module and providing a light-tight path. The low-sky background in the UV motivates attempts to limit the sensor dark current, and so the side-access panel becomes a radiator with an associated thermal path to the sensor.
In this way, our simple, quick, and easy camera in a box became a (tiny) UV telescope.

\subsubsection{What kind of telescope?}
With the available space on the side-panel adjacent to the free volume (40 mm) and the minimum vertical height occupied by the populated LUVCam PCBs + sensor ($\sim20$ mm), this leaves the remaining $\sim20$ mm for the entrance aperture of a telescope. We note that this aperture is smaller than the sensor itself! In order to baseline the optical design requirements, which (as above) will flow down to the rest of the system, we briefly explore the distribution of sources and their brightness in the UV sky.

The first (and last) survey of bright stars ($<8$ UV mag) in the UV sky was performed by ESRO's Ultraviolet Sky-Survey Telescope (aka S 2/68, \cite{td1a}) onboard the TD-1A spacecraft in 1973. This surprising fact can be explained by the realization that, due to the historically low UV sensitivity of silicon devices, the majority of UV telescopes on orbit over the last 50 years (including GALEX\cite{galex}, the UV surveyor) used Micro-Channel Plate (MCP) detectors. These detectors suffer damage from bright sources due to localized gain loss in the MCP, and degradation of the photocathode due to ion feedback from the MCP pores. For this reason, these telescopes purposefully avoided observations of bright stars, with GALEX having bright object limits of m$_{AB}$=9.5 and m$_{AB}$=8.9 in the far UV and near UV, respectively.

S 2/68 was a 27 cm off-axis telescope feeding spectrometer and photometer channels. The photometer channel was centered at 274 nm, with a bandwidth FWHM of 31 nm. It performed a drift scan survey of the entire sky with a cadence of $\sim6$ months, achieving a single-pass limiting visual magnitude of $\sim9$ for a B star. This relatively low sensitivity was caused by low total throughput of the optical system, the photocathode, and the photo-multiplier detectors. The resultant final system efficiency of $\sim1 \%$, corresponds to an effective area of $\sim5$ cm$^2$. In principle a $\sim20$ mm aperture, high-throughput UV telescope with modern coatings and our high UV QE sensor could approach the sensitivity of this much larger antique system.

\begin{figure}[ht]
    \centering
    \includegraphics[width=0.8\linewidth]{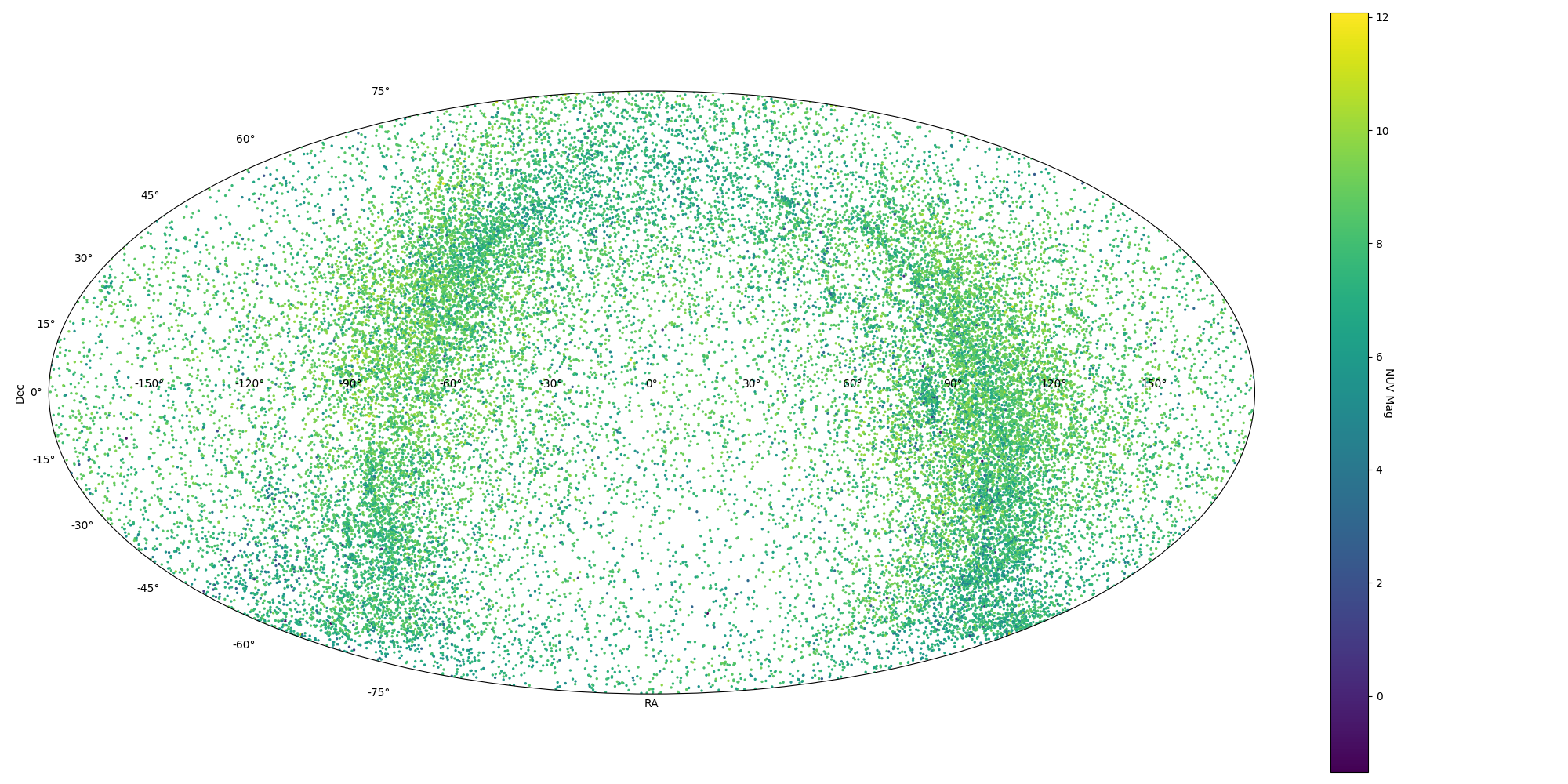}
    \caption{The sky distribution of sources in the TD1 Stellar Ultraviolet Fluxes catalog.}
    \label{fig:td1cat}
\end{figure}

The TD1 Catalog of Stellar Ultraviolet Fluxes \cite{td1catalog} contains measurements of 31,215 stars, down to $\sim12$ mag$_{AB}$ in the 274 nm (NUV) filter. Figure \ref{fig:td1cat} shows the distribution of these sources on the sky and as a function of magnitude. As can be seen, even in the Galactic Center, the source density is $<3$ stars per deg$^2$, and significantly more sparse at high galactic latitudes. This has significant implications for our optical design. 

While GRBBeta does have an ADCS to provide sensing, attitude determination, and 3-axis pointing control, this is a new and experimental system that is not required to work for GRBBeta mission success. As such, neither target acquisition, nor stable pointing, are guaranteed. Our telescope will therefore need to be able to blindly determine its own astrometric solution from a short exposure image, while possibly tumbling. Plate solving requires detecting many asterisms in the image, and comparing them to a catalog to uniquely identify the sky position, and orientation. \footnote{The sparsity of bright UV stars on the sky actually reduces the complexity of this task, as fewer asterisms are required to uniquely identify a given field, and break degeneracies.} A minimum number of sources is required to guarantee a solution.  Since the telescope aperture is bounded by the available area on the side panel, and the exposure time is necessarily short (in the case of unstable attitude), increasing the field-of-view (FOV) to be large enough to guarantee sufficient stars in each image is the appropriate path. Note that the allowed focal length of the telescope is bounded from above at $<75$ mm due to the geometry of the free volume, payload interface constraints, and size of the sensor. In the best case scenario, we wish to be able to take images at the highest possible resolution to maximize image quality and scientific return. Since the spacecraft is small, and the ADCS system a significant fraction of the total spacecraft inertial mass, the jitter induced by the reaction wheels into the spacecraft at high-frequencies will degrade the image quality irreducibly. A natural limit for the telescope resolution is therefore that the sampled PSF size be subdominant compared to the spacecraft jitter, thus bounding the allowed telescope focal length from below. In this way the bandpass (NUV; 250-300 nm), sensitivity (maximum achievable), field of view ($>7$\degree diameter), and resolution ($<1\sigma$ jitter over 1 second, 0.02\degree) design targets are set.

\subsection{Concept of Operations}
The fundamental requirements of the mission necessitate operating the sensor on orbit, and characterizing its performance as a function of orbital lifetime. Effectively, we wish to repeat much of the sensor characterization that was performed on the ground in Sec. \ref{sec:characterization}, [\citenum{sarik}], and [\citenum{ajay}], without the benefit of stabilized optical benches and assorted equipment, and with limited telemetry. 

We aim to measure the following parameters, all over a baseline mission of 3 months:
\begin{multicols}{2}
\begin{enumerate}
    \item Pixel operability (hot pixel accumulation rate)
    \item Per-pixel gain
    \item Read noise
    \item Linearity
    \item Full Well Capacity
    \item Dark current (as a function of temperature)
    \item Relative sensitivity
\end{enumerate}
\end{multicols}

Items 1-5 can all be measured by constructing a photon-transfer curve, as in the ground characterization. This fundamentally requires a $\sim$stable bright light source, relatively uniform illumination of a large number of pixels, a large number of dark pixels, variable exposure time from bias-level to saturation, and on-board storage of several images. These requirements can be met by imaging the Earth near-UV albedo in orbital daylight as a flat-field source.\footnote{In principle the photon transfer curve can also be constructed from dark measurements, but the pixel-to-pixel dark current non-uniformity is typically larger than pixel-to-pixel sensitivity non-uniformity, and additionally the exposure time to reach saturation for dark measurements will exceed any reasonable condition on sensor stability.} Based on ground testing, 3 images per exposure time, at 20 exposure times logarithmically spaced from bias until saturation, will produce an acceptable photon transfer curve. The relative brightness of the Earth albedo will allow us to reach saturation in short exposures, preventing significant temperature drift during measurement.

Dark current characterization requires a large number of optically dark pixels, accurate measurement of the sensor die temperature, and some thermal control. Based on ground tests, 10 images per exposure time, at 6 exposure times linearly spaced from 0.01 seconds to 60 seconds, should allow $<5\%$ precision measurements of the dark current. This necessitates maintaining the sensor temperature to within 2\degree C during at least 60 seconds of operation.

Quantum efficiency of the sensor cannot be directly measured, due to packaging constraints preventing integration of a NIST-traceable photodiode, and funding/time constraints preventing the study of photodiode stability in the orbital environment. Instead we will monitor the total system sensitivity on UV spectrophotometric standard stars, and can put upper limits on the sensor QE degradation by marginalizing out the uncertainty in the evolving throughput of the telescope optics (from contamination, etc). This necessitates sufficient sensitivity to achieve high SNR measurements of stellar flux, relatively stable pointing, imaging in eclipse, and precise knowledge of the pre-launch telescope throughput. 

All of these measurements will be repeated regularly, at a cadence set by the power and telemetry budget of the GRBBeta spacecraft, up to the end of the mission. Throughout operations single-event effects and other transient radiation effects on the camera control electronics will be monitored and recorded. Many of these measurements require the stable attitude control for up to 10s of seconds, and are therefore dependent on functionality of the experimental ADCS.

\subsection{Optical}

\begin{table}[h!]
\resizebox{\textwidth}{!}{%
\begin{tabular}{lll}
{\color[HTML]{1F497D} \textbf{ID}} & {\color[HTML]{1F497D} \textbf{Name}}           & {\color[HTML]{1F497D} \textbf{Requirement Text}} \\ \hline
\multicolumn{1}{|l|}{OPT-R-010}    & \multicolumn{1}{l|}{Clear Aperture}            & \multicolumn{1}{l|}{LUVCam Optics shall provide $>15$ mm entrance pupil diameter optical path from sensor to space along +Y direction} \\ \hline
\multicolumn{1}{|l|}{OPT-R-020}    & \multicolumn{1}{l|}{Point Spread Function}     & \multicolumn{1}{l|}{LUVCam Optics shall provide PSF 1-sigma to \textless 1/2  1 sigma jitter spec at all field angles.}             \\ \hline
\multicolumn{1}{|l|}{OPT-R-030}    & \multicolumn{1}{l|}{Throughput}                & \multicolumn{1}{l|}{LUVCam Optics shall provide \textgreater{}80\% throughput (240-310 nm) along optical path.}                     \\ \hline
\multicolumn{1}{|l|}{OPT-R-040}    & \multicolumn{1}{l|}{Scattered and Stray Light} & \multicolumn{1}{l|}{LUVCam Optics shall mitigate scattered + stray light to level of sky background.}                                      \\ \hline
\multicolumn{1}{|l|}{OPT-R-050}    & \multicolumn{1}{l|}{Field of View}             & \multicolumn{1}{l|}{LUVCam Optics shall provide unvignetted image over 1k x 1k pixels.}                                             \\ \hline
\multicolumn{1}{|l|}{OPT-R-060}   & \multicolumn{1}{l|}{Red Leak}                  & \multicolumn{1}{l|}{LUVCam optics shall attenuate light \textgreater{}310 nm from reaching sensor to \textgreater{}OD3.5.}          \\ \hline
\multicolumn{1}{|l|}{OPT-R-070}    & \multicolumn{1}{l|}{Plate Scale}               & \multicolumn{1}{l|}{LUVCam optics shall provide a plate scale of no less than 25"/pix and no more than 35"/pix.}                    \\ \hline
\multicolumn{1}{|l|}{OPT-R-080}    & \multicolumn{1}{l|}{Separation}                & \multicolumn{1}{l|}{LUVCam optics shall provide at least 1k x 1k dark pixels, having no intersection with image sub-frame.}                         \\ \hline
\end{tabular}%
}
\caption{Requirements for the LUVCam optical system.}
\label{tab:optical}
\end{table}

\begin{figure}[ht]
    \centering
    \includegraphics[width=0.7\textwidth]{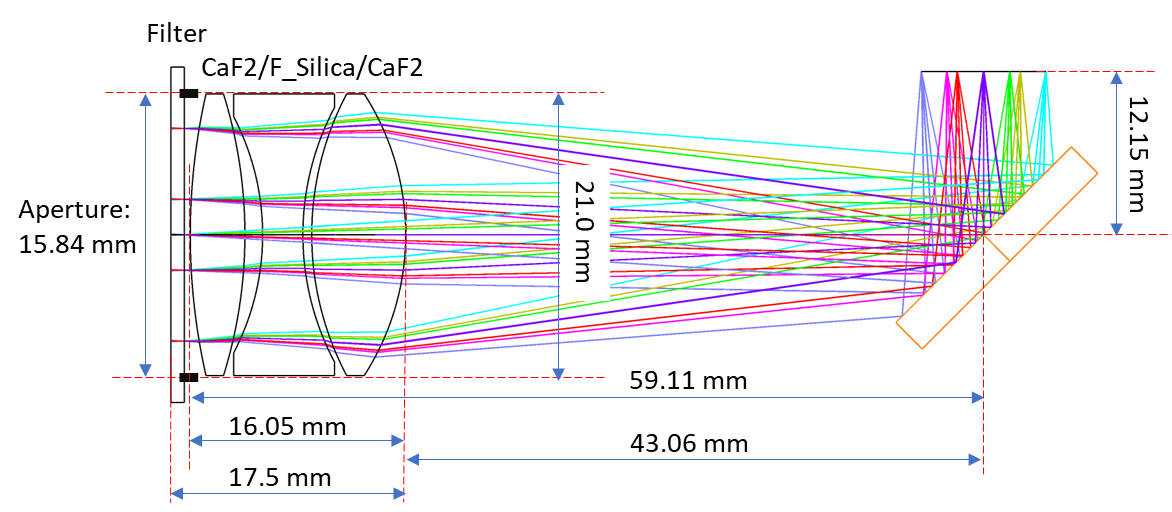}
    \caption[Optical diagram of the LUVCam mini telescope.]{Optical diagram of the LUVCam mini telescope, an aperture stop sits behind the filter ensuring an un-vignetted image over the 1000 pixel field.}
    \label{fig:optical-design}
\end{figure}

\begin{figure}[h!]
\centering
\begin{minipage}{.4\textwidth}
  \centering
  \begin{tabular}{cc}
    {\color[HTML]{1F497D} \textbf{Parameter}} & {\color[HTML]{1F497D} \textbf{Specification}} \\ \hline
    \multicolumn{1}{|c|}{Field of View}       & \multicolumn{1}{c|}{9.5 deg (diam)}          \\ \hline
    \multicolumn{1}{|c|}{Bandpass}            & \multicolumn{1}{c|}{240-310 nm}              \\ \hline
    \multicolumn{1}{|c|}{F/\#}                & \multicolumn{1}{c|}{3.5}                     \\ \hline
    \multicolumn{1}{|c|}{Focal Length}        & \multicolumn{1}{c|}{55.77 mm @ 250 nm}       \\ \hline
    \multicolumn{1}{|c|}{Spot Size}           & \multicolumn{1}{c|}{0.02 deg @ EE80}         \\ \hline
    \multicolumn{1}{|c|}{Throughput}          & \multicolumn{1}{c|}{$94 \pm 0.1$ \%}         \\ \hline
    \multicolumn{1}{|c|}{Distortion}          & \multicolumn{1}{c|}{\textless 0.01 \%}       \\ \hline
  \end{tabular}
  \captionof{table}{The LUVCam telescope final performance specifications.}
  \label{tab:telespecs}
\end{minipage}%
\hspace{1cm}
\begin{minipage}{.52\textwidth}
  \centering
  \includegraphics[width=\linewidth]{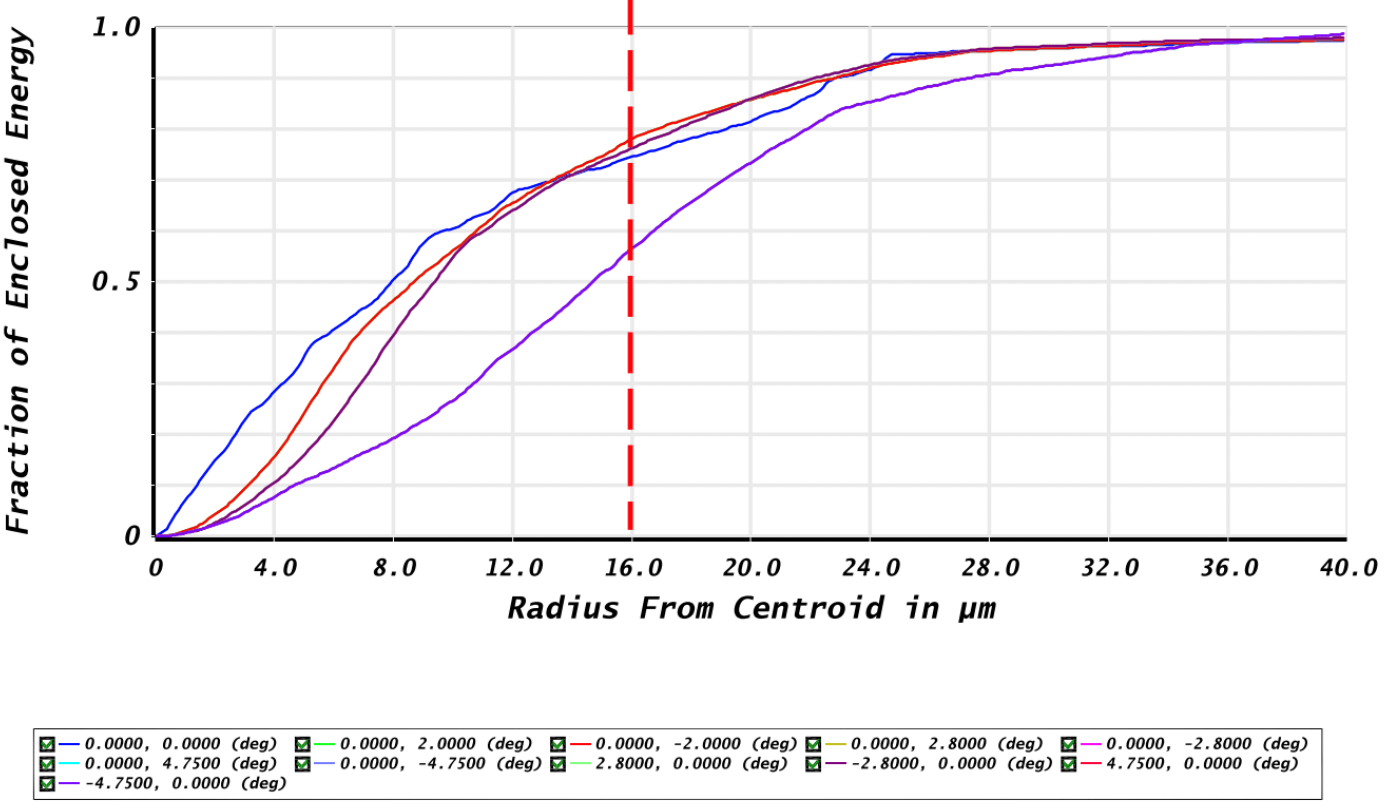}
  \caption{Enclosed energy as a function of radius from the spot centroid, for different field positions.}
  \label{fig:enclosed-energy}
\end{minipage}
\end{figure}

\begin{figure}
    \centering
    \includegraphics[width=0.5\linewidth]{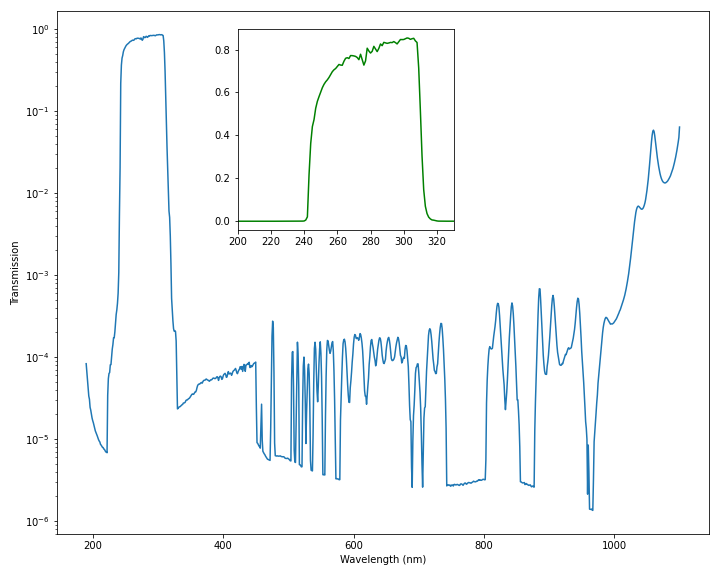}
    \caption[The absolute transmission of the LUVCam bandpass filter, flight article.]{Spectrometer measurement of the absolute transmission of the flight article Asahi bandpass filter, from 190-1100 nm.}
    \label{fig:filter}
\end{figure}

The optical requirements are met by a folded triplet apochromat design, shown in Fig. \ref{fig:optical-design}. This provides a 55 mm focal length, achieving a plate scale on the sensor of 34" per pixel. The design is optimized for light at 230-310 nm, and produces a PSF with an 80\% enclosed energy radius of $<2$ pixels, out to 4.5 degrees from the center of the field (see Fig. \ref{fig:enclosed-energy}). The refractive optics are custom designed by our team, and fabricated and coated by YFD Optics, with Ti$_3$O$_5$ and SiO$_2$ AR coatings producing $<0.5\%$ reflection at each surface. The fold mirror is a COTS product, a 20 mm diameter Fused Silica substrate with an Enhanced Aluminum coating, providing $>90\%$ reflectivity across the bandpass. The chosen filter is a COTS product from Asahi Spectra, a Fused Silica substrate coated with multiple layers of SiO$_2$ and HfO$_2$, producing a bandpass from 240-310 nm, and OD3.5 or better attenuation maintained from 310-1000 nm (see Fig. \ref{fig:filter}). The optics illuminate a subset of the sensor ($\sim$1k x 1k pixels), and the remaining pixels are kept dark by a light-tight enclosure painted with black Aeroglaze Z306, whose entrance serves as a square field stop. A short  ($<6$ mm) painted baffle sits in front of the optics, length limited by the CubeSat deployer keep-out zone.

\begin{table}[ht]
\resizebox{\textwidth}{!}{%
\begin{tabular}{ccccccc}
{\color[HTML]{1F497D} \textbf{Part No.}} & {\color[HTML]{1F497D} \textbf{Description}} & {\color[HTML]{1F497D} \textbf{Material}}        & {\color[HTML]{1F497D} \textbf{Coatings}} & {\color[HTML]{1F497D} \textbf{Clear Aperture Dia (mm)}} & {\color[HTML]{1F497D} \textbf{Tilt Tol. (+- deg)}} & {\color[HTML]{1F497D} \textbf{Decentre Tol. (+- mm)}} \\ \hline
\multicolumn{1}{|c|}{P173-1}             & \multicolumn{1}{c|}{Baffle}                 & \multicolumn{1}{c|}{Al 6061}                    & \multicolumn{1}{c|}{Aeroglaze Z306}      & \multicolumn{1}{c|}{16.99}                              & \multicolumn{1}{c|}{-}                             & \multicolumn{1}{c|}{0.1}                              \\ \hline
\multicolumn{1}{|c|}{C186-1}             & \multicolumn{1}{c|}{Filter}                 & \multicolumn{1}{c|}{Fused Silica, UV Grade}     & \multicolumn{1}{c|}{SiO2, HfO2}          & \multicolumn{1}{c|}{15.95}                              & \multicolumn{1}{c|}{1.0}                             & \multicolumn{1}{c|}{1.0}                                \\ \hline
\multicolumn{1}{|c|}{P172-1}             & \multicolumn{1}{c|}{Aperture Stop}          & \multicolumn{1}{c|}{CRES 316, Cond A}           & \multicolumn{1}{c|}{-}                   & \multicolumn{1}{c|}{15.84}                              & \multicolumn{1}{c|}{0.50}                          & \multicolumn{1}{c|}{0.05}                             \\ \hline
\multicolumn{1}{|c|}{P176-1}             & \multicolumn{1}{c|}{Lens 1}                 & \multicolumn{1}{c|}{Calcium Fluoride, UV Grade} & \multicolumn{1}{c|}{Ti3O5, SiO2}         & \multicolumn{1}{c|}{16.04}                              & \multicolumn{1}{c|}{0.05}                          & \multicolumn{1}{c|}{0.025}                            \\ \hline
\multicolumn{1}{|c|}{P177-1}             & \multicolumn{1}{c|}{Lens 2}                 & \multicolumn{1}{c|}{Fused Silica, UV Grade}     & \multicolumn{1}{c|}{Ti3O5, SiO2}         & \multicolumn{1}{c|}{16.04}                              & \multicolumn{1}{c|}{0.05}                          & \multicolumn{1}{c|}{0.025}                            \\ \hline
\multicolumn{1}{|c|}{P178-1}             & \multicolumn{1}{c|}{Lens 3}                 & \multicolumn{1}{c|}{Calcium Fluoride, UV Grade} & \multicolumn{1}{c|}{Ti3O5, SiO2}         & \multicolumn{1}{c|}{18.23}                              & \multicolumn{1}{c|}{0.10}                          & \multicolumn{1}{c|}{0.050}                            \\ \hline
\multicolumn{1}{|c|}{P187-1}             & \multicolumn{1}{c|}{Mirror}                 & \multicolumn{1}{c|}{Fused Silica}               & \multicolumn{1}{c|}{Enhanced Al.}        & \multicolumn{1}{c|}{15.92}                              & \multicolumn{1}{c|}{0.10}                          & \multicolumn{1}{c|}{1.0}                              \\ \hline
\multicolumn{1}{|c|}{P168-1}             & \multicolumn{1}{c|}{Field Stop}             & \multicolumn{1}{c|}{GF30 PEEK}                  & \multicolumn{1}{c|}{Aeroglaze Z306}      & \multicolumn{1}{c|}{12.59}                               & \multicolumn{1}{c|}{-}                             & \multicolumn{1}{c|}{2.0}                                \\ \hline
\end{tabular}%
}
\caption{The LUVCam telescope optical components, materials, coatings, aperture, and alignment tolerances.}
\label{tab:opticalchain}
\end{table}

\subsection{Thermal}

\begin{table}[ht]
\resizebox{\textwidth}{!}{%
\begin{tabular}{lll}
{\color[HTML]{1F497D} \textbf{ID}} & {\color[HTML]{1F497D} \textbf{Name}}           & {\color[HTML]{1F497D} \textbf{Requirement Text}} \\ \hline
\multicolumn{1}{|l|}{THRM-R-010}    & \multicolumn{1}{l|}{Sensor Temperature Control}            & \multicolumn{1}{l|}{Shall provide sensor cooling to 0C during sensor operation.} \\ \hline
\multicolumn{1}{|l|}{THRM-R-020}    & \multicolumn{1}{l|}{Thermal Interfaces+Path}     & \multicolumn{1}{l|}{Shall provide thermal interfaces on sensor, and path to radiator surface}             \\ \hline
\multicolumn{1}{|l|}{THRM-R-030}    & \multicolumn{1}{l|}{Radiator Area}                & \multicolumn{1}{l|}{Radiator shall not exceed free exterior area as defined by ICD, shall provide access slot to OBC.}                     \\ \hline
\end{tabular}%
}
\caption{Requirements for the LUVCam passive thermal control system.}
\label{tab:thermal}
\end{table}

Characterizing the effect of the orbital environment (predominantly radiation) on the performance of the sensor as a function of temperature is a critical measurement. In addition, the low sky background in the UV motivates low dark current operation. Both of these necessitate a thermal control system capable of significant heat rejection; the sensor itself dissipates 1400 mW during operation, and the control electronics (thermally isolated from the sensor to the extent feasible) another $\sim3300$ mW. Heat rejection is a difficult task with limited volume and exterior area to work with, and a spacecraft that was not designed with any thermal control in mind. Actually achieving background limited observations ($\sim 0.02$ ct/s/pix) is not feasible due to the expected short exposure times and sub-optimal thermal environment of the small spacecraft. A modest target operating temperature of 0\degree C was chosen, which  ensures that the system remains read noise limited for exposure times up to 10 seconds.
\begin{figure}[h!]
    \centering
    \includegraphics[width=0.5\linewidth]{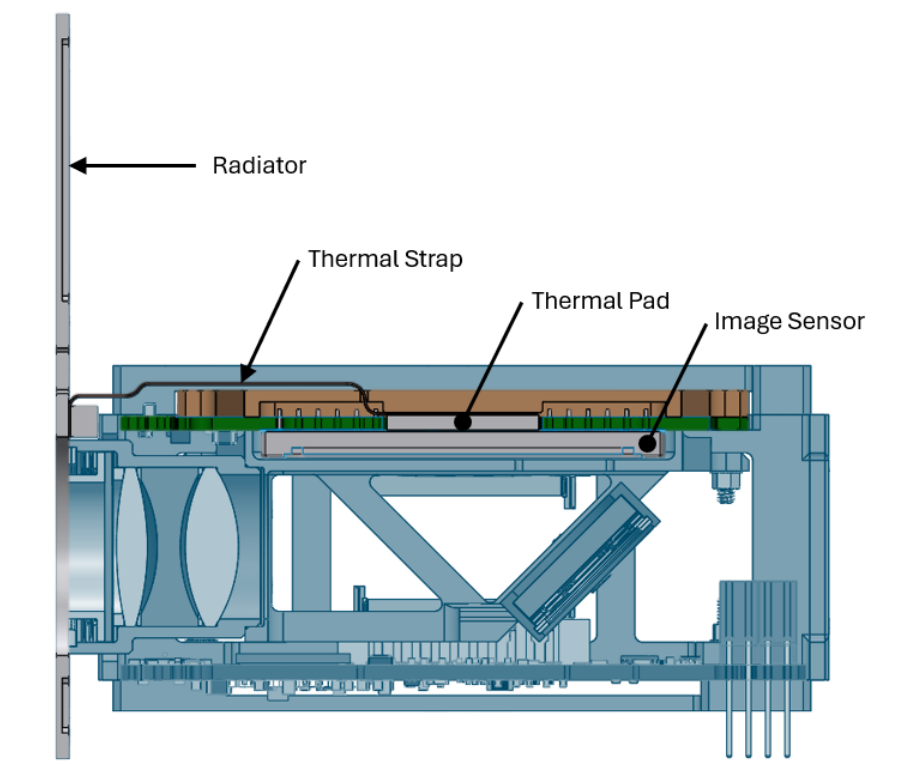}
    \caption[Cross-section of the LUVCam passive thermal control system.]{The thermal path from the sensor to the radiator. The radiator is thermally isolated from the rest of the payload, and spacecraft, by PEEK fasteners and plastic spacers (not shown).}
    \label{fig:thermal-path}
\end{figure}

A simple passive thermal control system was designed, substituting the -X+Z exterior panel for a custom radiator with a short thermal strap to the sensor. The thermal path from the sensor to deep space is shown in Fig. \ref{fig:thermal-path}. The backside of the sensor package is prepared with a thin layer of Apiezon L grease. A copper thermal strap is sandwiched between an aluminum thermal pad contacting the backside of the sensor (also coated with Apiezon L) and the back-side detector enclosure. This thermal strap extends out of the sensor enclosure via a narrow feed, approximately $\sim4$ cm to where it is fastened to the radiator by another Apiezon shmear sandwich. The radiator plate is made of clear anodized aluminum, and is thermally isolated from the rest of the spacecraft via a FR-4 spacer and GF30 PEEK/steel fasteners. The plate provides $>70$ cm$^2$ of radiating surface area.

\begin{figure}[ht]
\centering
    \includegraphics[width=0.75\linewidth]{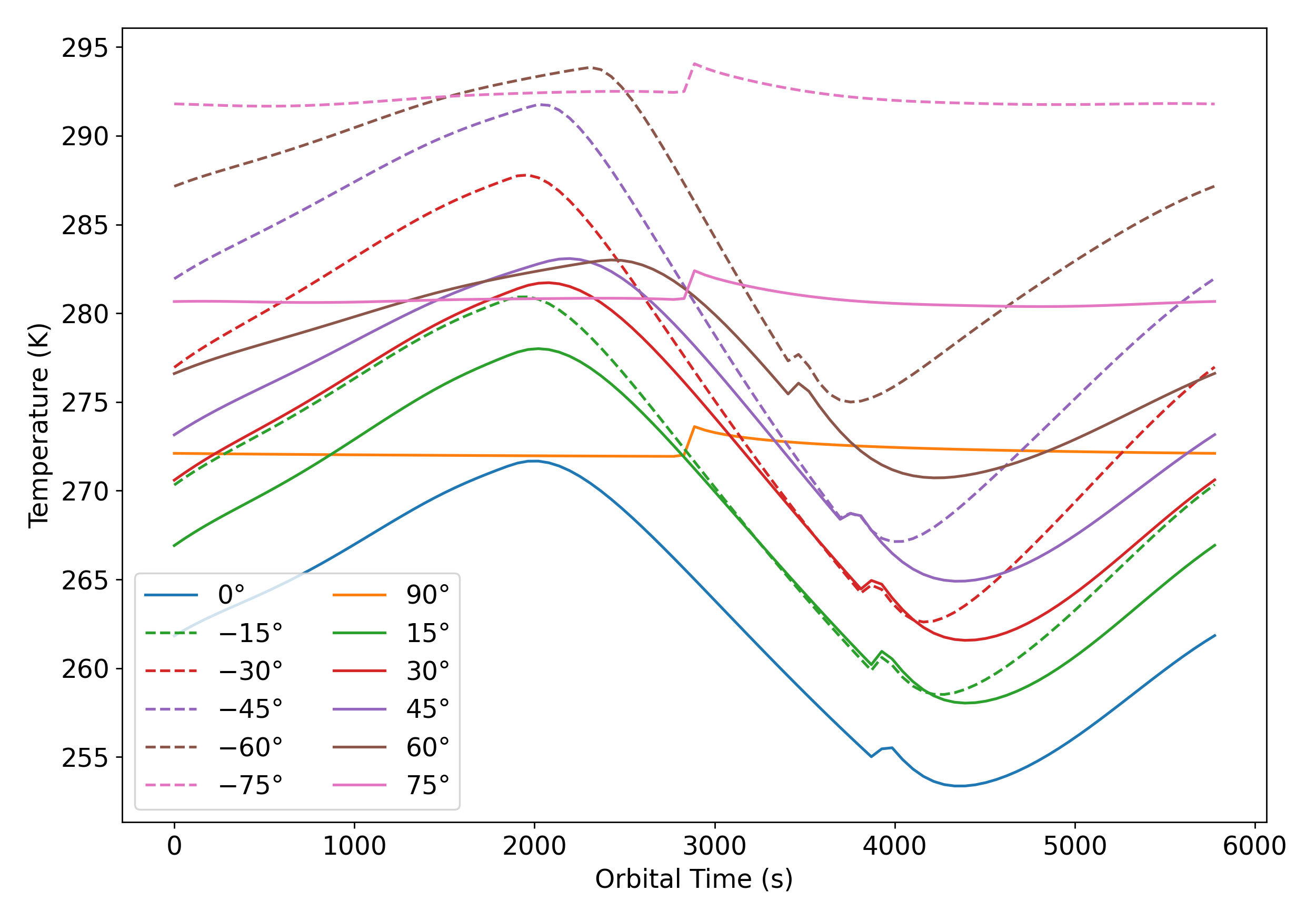}
    \caption[Modelled LUVCam sensor temperature as a function of beta angle and orbital time.]{Thermal modelling of the sensor temperature over a single orbit with varying beta angle. The sensor is activated for imaging for the last 60 seconds of eclipse. For non-negative beta angles the ADCS is active and maintaining the radiator pointing at deep space. For negative beta angles the radiator pointing is allowed to drift over the Sun and Earth, to explore the impact of a non-functional ADCS.}
    \label{fig:temp-profile}
\end{figure}

To predict the thermal performance of the system, we constructed a detailed finite element thermal model of the LUVCam payload and the GRBBeta spacecraft, with its primary power producing elements. This model includes all appropriate material thermal properties, optical properties, conductance of interfaces, and heat loads. The orbital environment at 580 km with inclination of 62\degree\ is input into the model, which includes time resolved solar heat flux, Earth IR heat flux, Earth albedo, and time-dependent spacecraft orientation and power profiles. In operation, we will keep the radiator pointing at deep space, and away from the Sun and Earth at all times. For the purpose of studying the impact of a non-functional ADCS system, we also model cases where the radiator pointing cannot be optimally maintained. Representative results of the sensor temperature at various beta angles is shown in \ref{fig:temp-profile}, with the sensor powered on for 60 s near the end of eclipse. This shows that sensor temperatures below 0\degree C can be achieved for short imaging operations across a range of beta angles and ADCS functionality.

\begin{figure}
    \centering
    \includegraphics[width=0.6\linewidth]{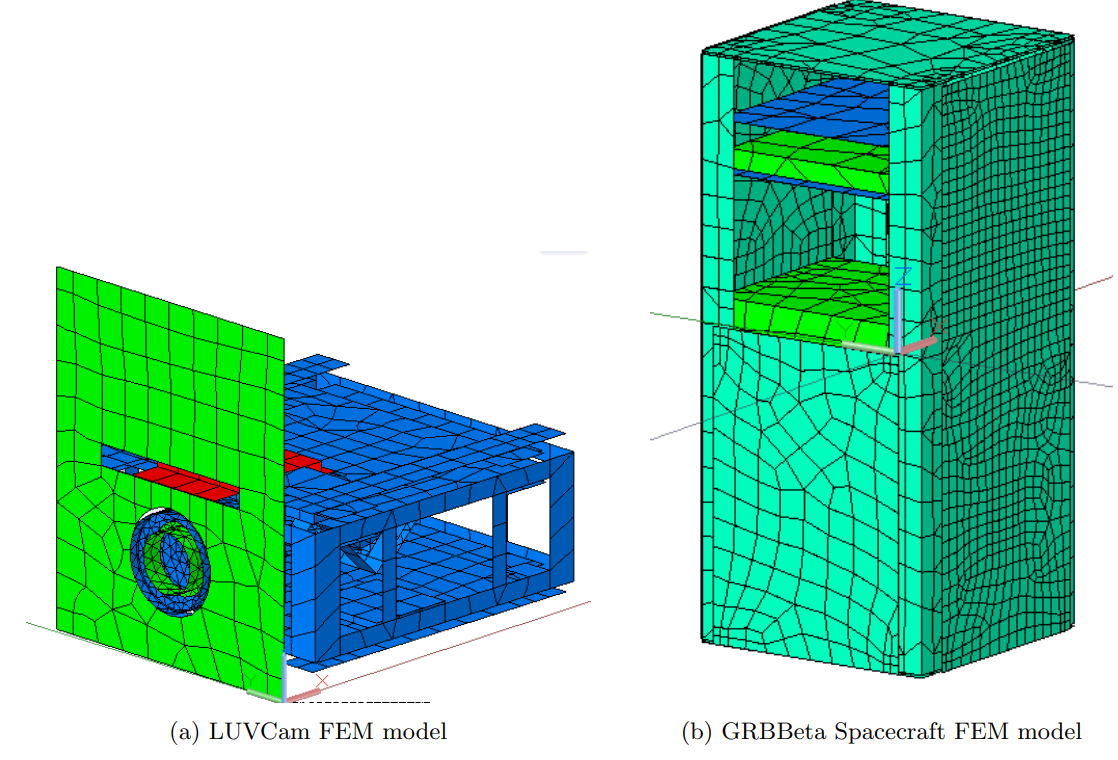}
    \caption[GRBBeta and LUVCam finite-element model.]{The finite element model mesh of the LUVCam payload and GRBBeta spacecraft.}
    \label{fig:fem-model}
\end{figure}

\subsection{Payload}
The complete LUVCam Tech Demo payload is 96 x 96 x 44.5 mm$^3$ with a total mass of 287.3 grams. It comprises a few dozen custom designed optomechanical components, machined out of GF30 PEEK for mass savings where manufacturing tolerance, thermal, and other demands allowed, otherwise Al 6061-T6, and CRES 304 stainless steel were used. The sensor and control electronics subsystem (2 custom PCBs, connectors, sensor, flex cable) comprises 131.7 grams, the optics subsystem (lenses, mirrors, spacers, stops, springs, filter, baffle, barrel assembly, support) comprises 27.5 grams, the thermal subsystem (pad, strap, radiator, retainer) comprises 34.3 grams, and the structural elements (sensor enclosure, sensor support, central frame, interface adapters, fasteners, tape, etc) contribute $< 90$ grams. 
\begin{figure}[ht]
    \centering
    \includegraphics[width=0.8\linewidth]{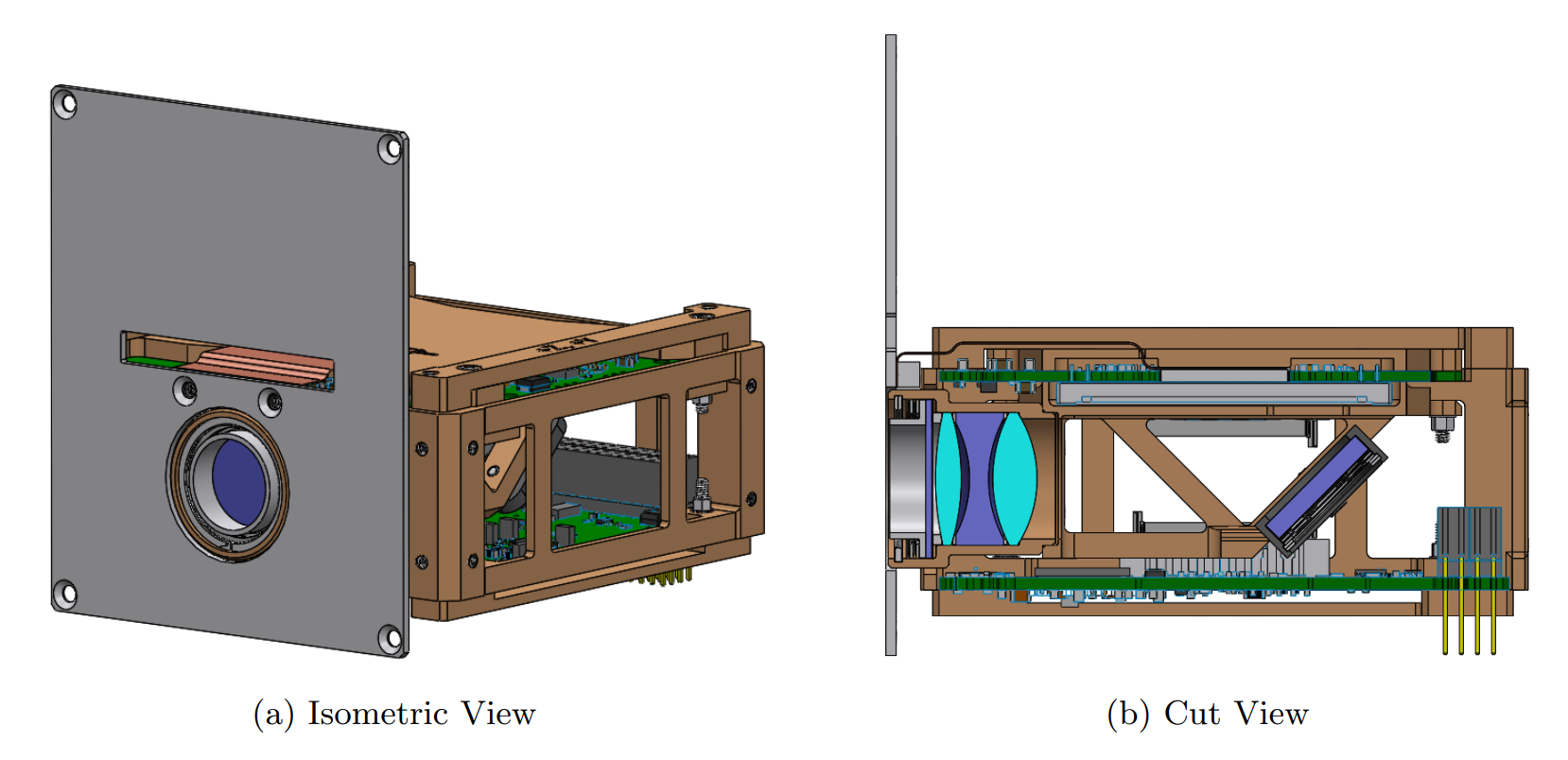}
    \caption[Render of the complete LUVCam Tech Demo payload.]{The LUVCam Tech Demo payload. Flex connectors and light-tight baffling around optics support enclosure are not shown in this model.}
    \label{fig:payload}
\end{figure}

All components (except for the image sensor) were thoroughly cleaned and vacuum baked to accelerate out-gassing prior to assembly. After an initial 24 hour low-pressure (1E-3 Torr), high temperature (140$^\circ$C) bake-out, components were placed in the LUVCam thermal-vacuum chamber (3E-6 Torr average) and heated to 80$^\circ$C (significantly above temperatures expected on orbit) for 24 hours, while an optical witness sample in the form of a fused silica window was kept at -10$^\circ$C. These windows were inspected, and their transmission was measured, before and after the bake-out contamination test, with $<<1 \%$ total transmission loss measured across the 240-310 nm bandpass. 

\begin{figure}
    \centering
    \subfloat{\includegraphics[width=0.33\linewidth]{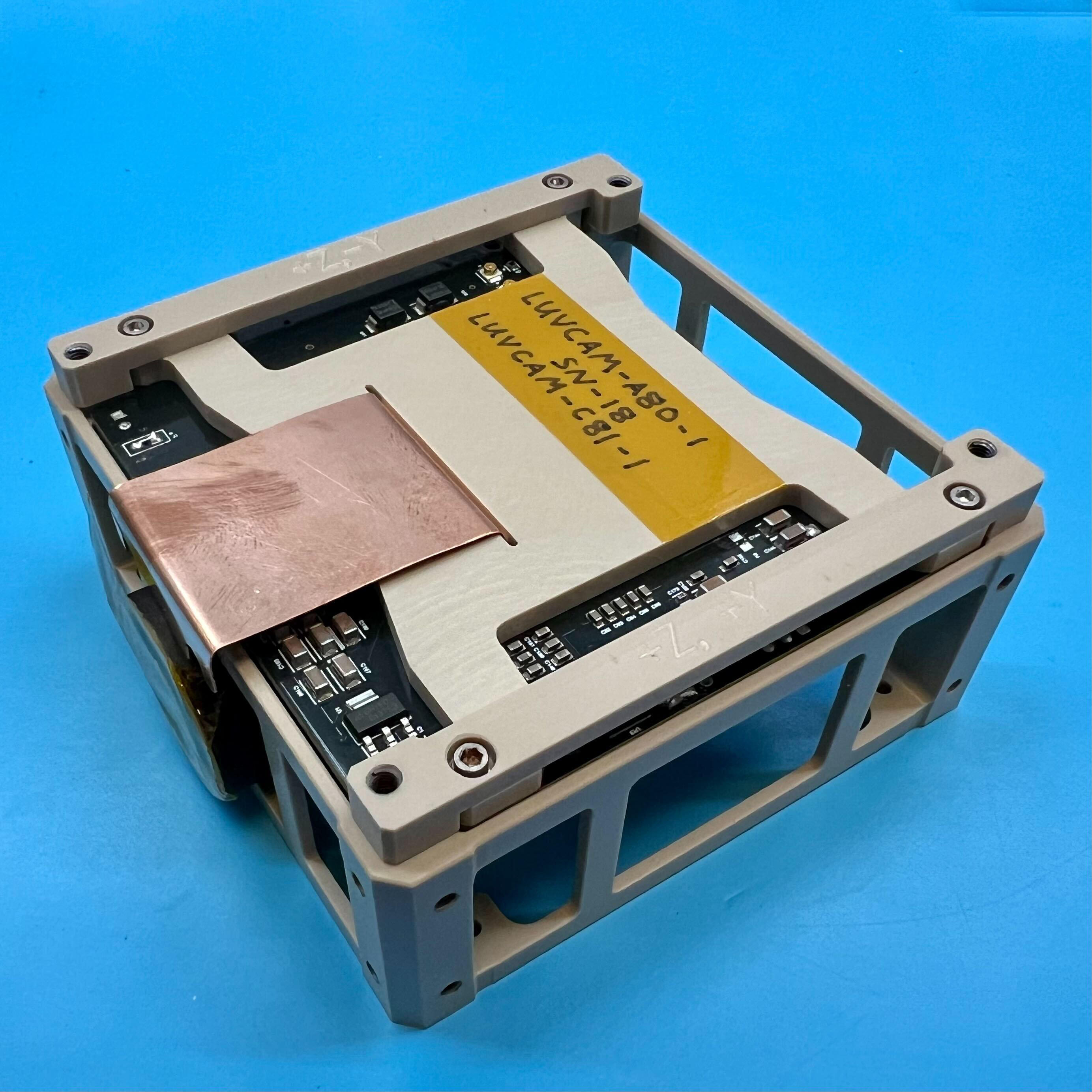}}
    \subfloat{\includegraphics[width=0.33\linewidth]{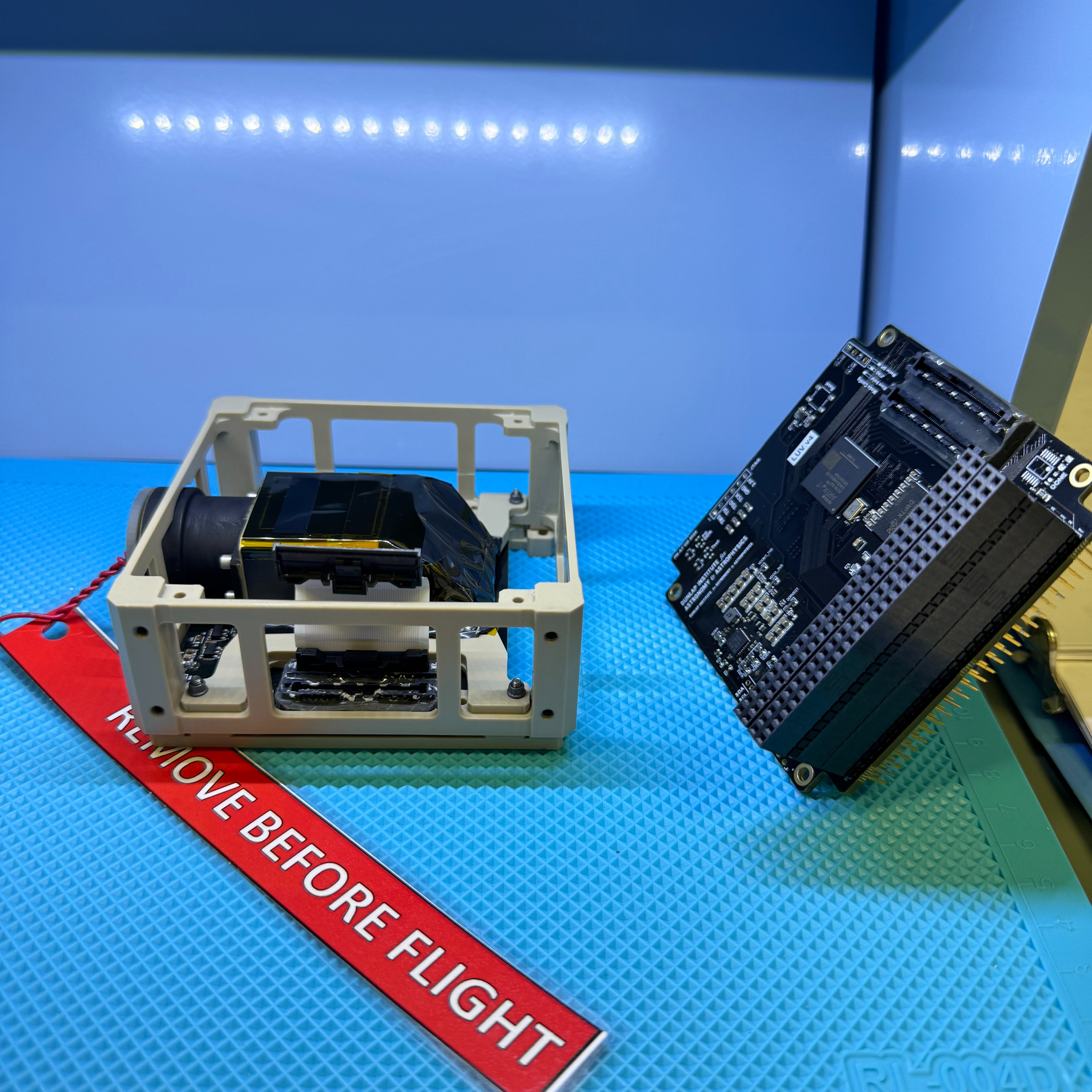}}
    \subfloat{\includegraphics[width=0.33\linewidth]{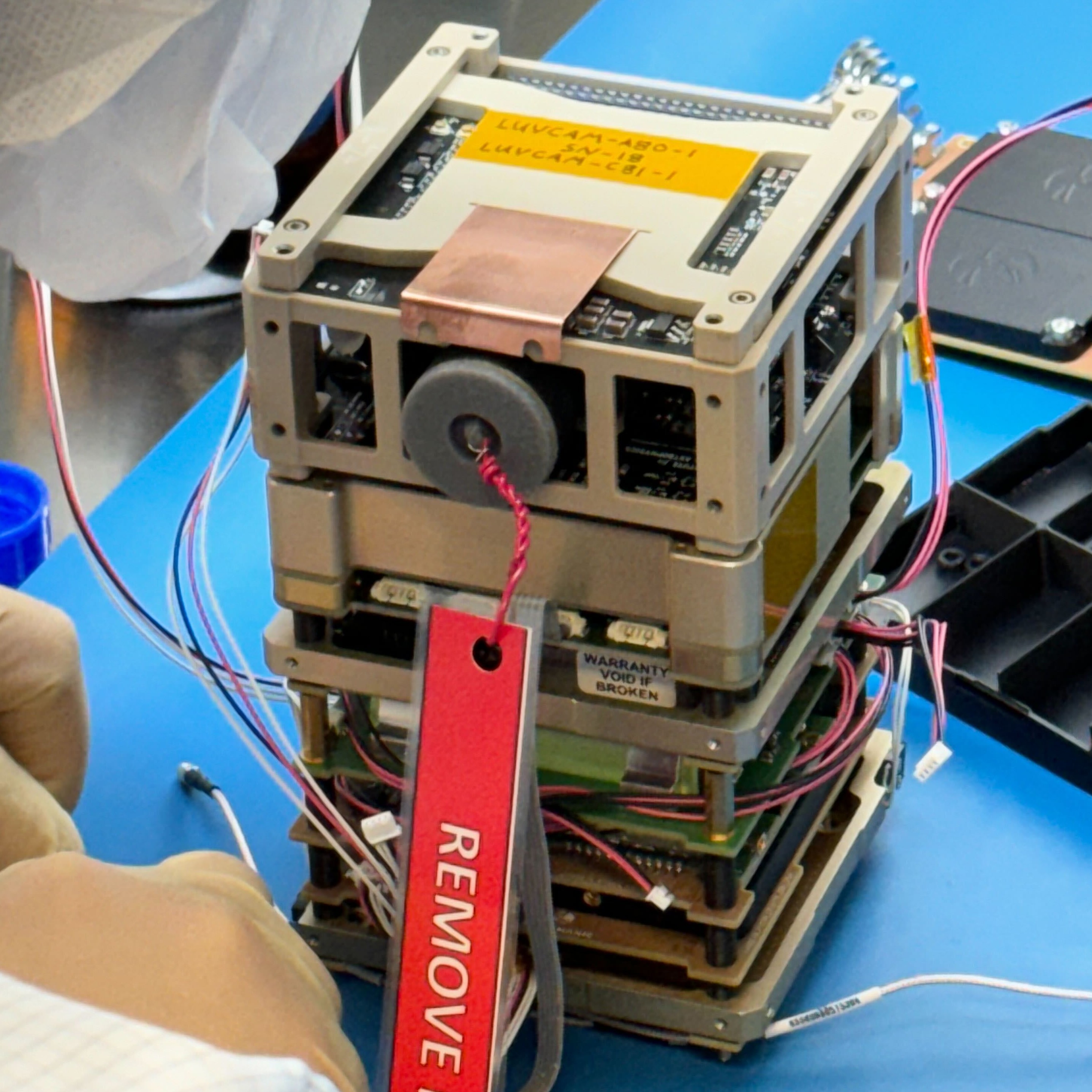}}
    \caption[The LUVCam Tech Demo Payload during assembly and integration.]{\textit{Left:} The LUVCam payload +Z side, showing the thermal strap, sensor PCB, and backside of the sensor enclosure, prior to control PCB integration. \textit{Center:} LUVCAM payload internals (left), showing sealed optical system during integration of control PCB (right) and installation of flex connectors. \textit{Right:} The LUVCam payload being integrated to the GRBBeta stack.}
    \label{fig:payload-assemblies}
\end{figure}

\begin{figure}
    \centering
    \includegraphics[width=0.75\linewidth]{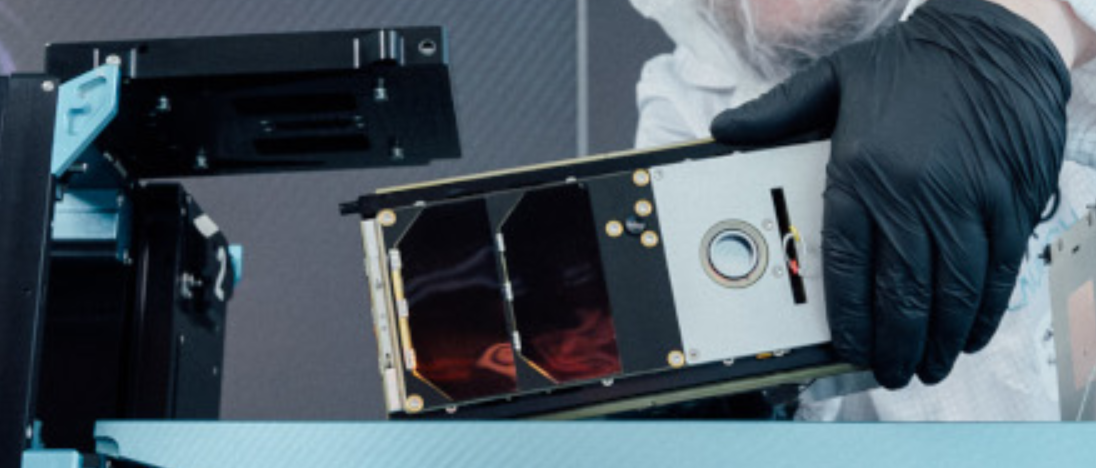}
    \caption[LUVCam Tech Demo in GRBBeta being placed into the CubeSat deployer.]{The GRBBeta spacecraft being placed into the CubeSat deployer. The LUVCam optics and radiator can be seen, just after removal of the protective lens cap. Image credit: ExoLaunch}
    \label{fig:deployer}
\end{figure}

For this first flight, the Front Side Illuminated (FSI) version of the sensor was integrated to the camera. Assembly was performed in an ISO-5 laminar flow bench at the Dunlap Institute, where the sensor and optics were sealed, before the final integration of the LUVCam payload to the GRBBeta spacecraft at Masaryk University in Brno, Czechia (Fig. \ref{fig:payload-assemblies}). Following final integration, the GRBBeta spacecraft underwent ADCS and gamma-ray detector calibration activies at VZLU in Prague, before being delivered to Exolaunch in Berlin for integration with its deployer (Fig. \ref{fig:deployer}). The protective lens cap was removed from LUVCam just prior to final insertion into the deployer. GRBBeta was launched from Guiana Space Center on July 9, 2024, on the inaugural flight of the Ariane 6 rocket, and deployed into a 579 km orbit. The GRBBeta spacecraft is working well, and LUVCam characterization is in progress.

\begin{figure}[ht]
    \centering
    \subfloat{\includegraphics[width=0.5\textwidth]{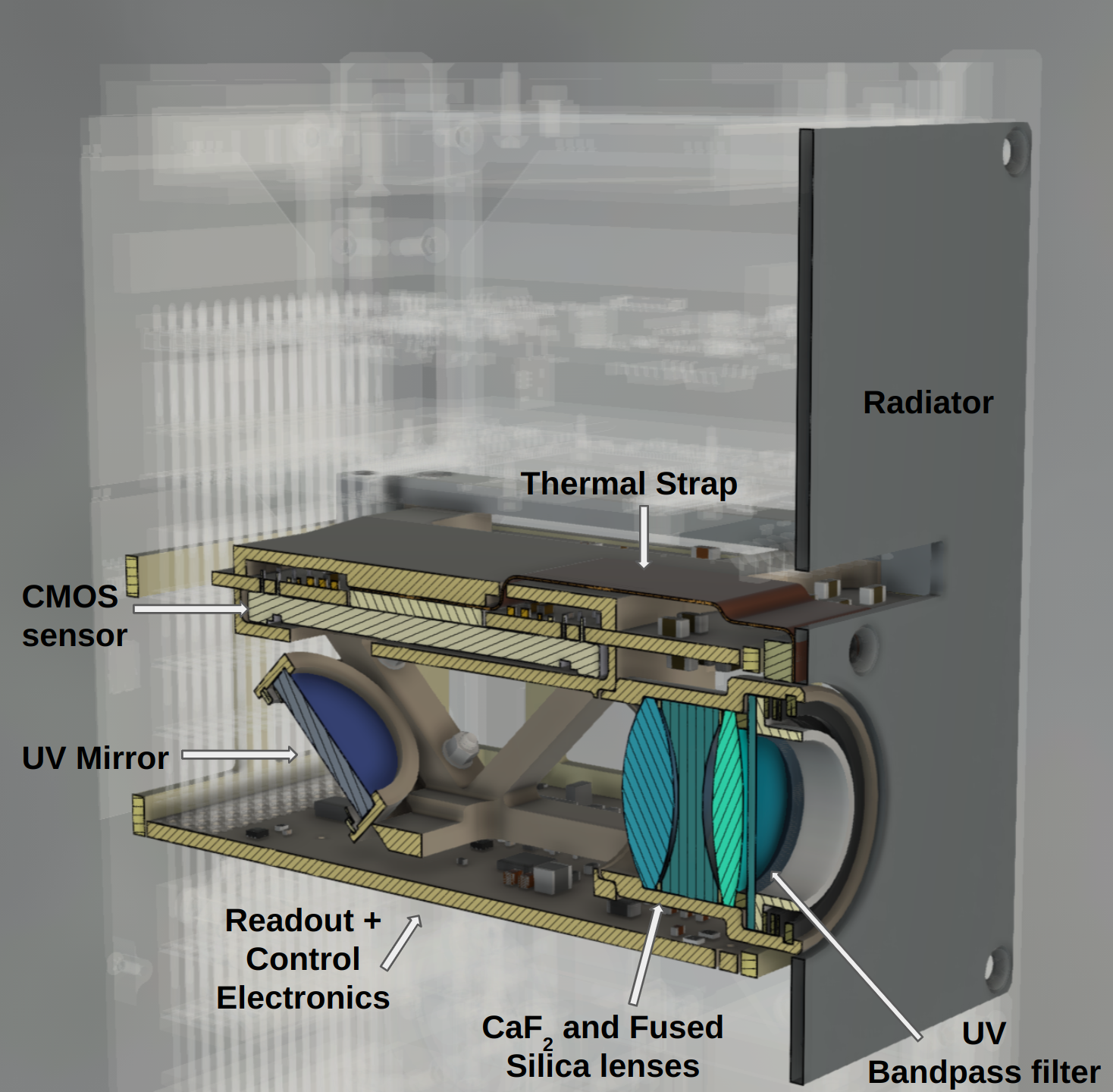}}
    \subfloat{\includegraphics[width=0.491\textwidth]{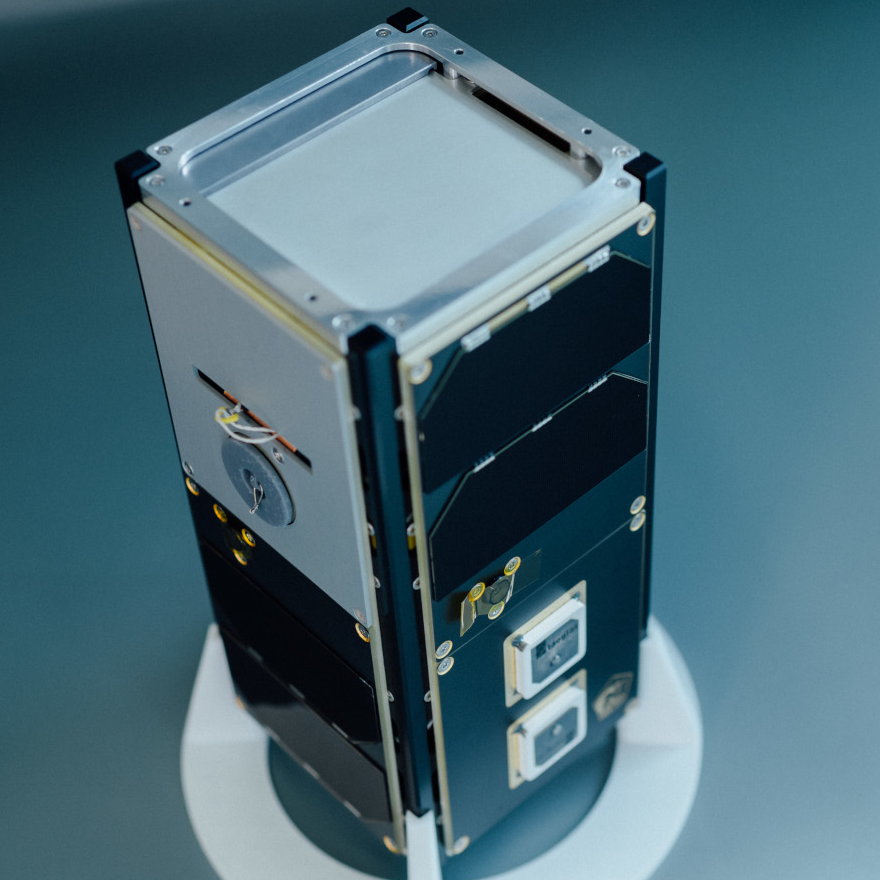}}
    \caption[The LUVCam Tech Demo inside GRBBeta.]{\textit{Left:} Cross-section of the LUVCam payload inside GRBBeta spacecraft. Flex cables and light-tight soft baffling around the optical path are not shown. \textit{Right:} The fully assembled GRBBeta, with RBF lens cap over LUVCam optics.}
    \label{fig:GRBBetacut}
\end{figure}

\subsection{Radiation Environment}
A key goal of the technology demonstration mission is to characterize the sensor degradation as a function of orbital lifetime, and particularly the effects of radiation damage. Understanding the radiation environment is therefore critical to interpreting the measurement results. GRBBeta's 580 km, 62 degree inclination orbit takes it through both the South Atlantic Anomaly and the Auroral/polar regions, experiencing the trapped energetic protons and electrons in addition to the background radiative flux from cosmic rays and the solar wind. In this environment, the bare sensor would accumulate $>100$ krad TID over 1 year in orbit.

The sensor is sealed on all sides by a GF30 PEEK enclosure with minimum thickness of $\sim1.5$ mm, with a small opening for the optical path (see Fig. \ref{fig:GRBBetacut} left, and Fig. \ref{fig:payload} right). However, the surrounding optical and thermo-mechanical structure of LUVCam, as well as the  superstructure and other payloads of the GRBBeta spacecraft, result in a highly non-isotropic shielding profile. In lieu of complex radiation transport modelling at the component level, we ray-trace the GRBBeta spacecraft, sampling the Al-equivalent column density and number of interaction layers, along various lines of sight. This approach is a reasonable approximation to first order \cite{carefulcots}. The sensor has $>3$ mm Al-equivalent shielding along all directions except for a pathway that provides direct passage from space to the rear of the sensor enclosure via the OBC access port in the radiator panel. However, these lines-of-sight make up a very small solid angle, $<2$\% of sphere, and still retain $>2$ mm Al-equivalent shielding. We therefore conservatively allocate 200 rad over 1 year to this solid angle. Along other sight-lines most material interactions are with GF30 PEEK, PCB and assorted components, Aluminum, Cesium Iodide, plastic polymers, PV panels, and others. Subsequently we apply a weighted average over the sky to determine the average number of interaction layers and Aluminum equivalent thickness thereof. 

We employed the Space Environment, Effects, and Education System (SPENVIS) simulation tool and the NASA AE-8/AP-8 model (at solar maximum) of electron and proton fluxes, coupled with the International Geomagnetic Reference Field (IGRF). SPENVIS provides access to the SHIELDOSE-2 model, to which we input our equivalent shielding parameters with a Silicon target, over a year-long mission with GRBBeta's orbital elements. We calculate a total ionizing dose of $<1.7$ krad after 1 year. This calculation includes input trapped electrons and protons, solar protons, and bremsstrahlung produced by shield interactions. Higher precision modelling is left for future work. We note that the GRBBeta gamma-ray detector provides a useful in-situ companion tracer for the local environment of photons and charged particles with energies from 70-890 keV.

\subsection{Expected Performance}
\begin{figure}[ht]
    \centering
    \subfloat{\includegraphics[width=0.5\linewidth]{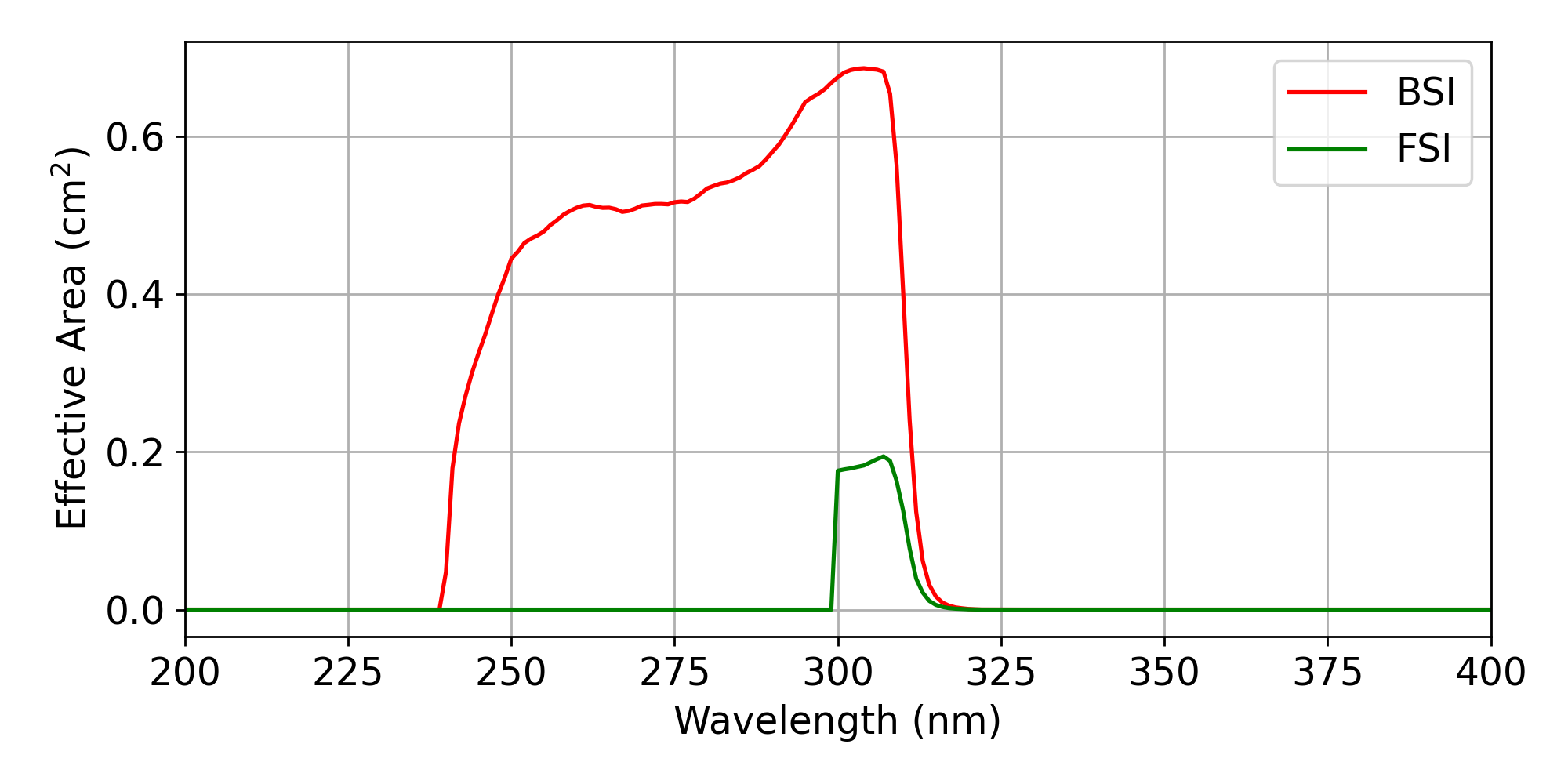}}
    \subfloat{\includegraphics[width=0.5\linewidth]{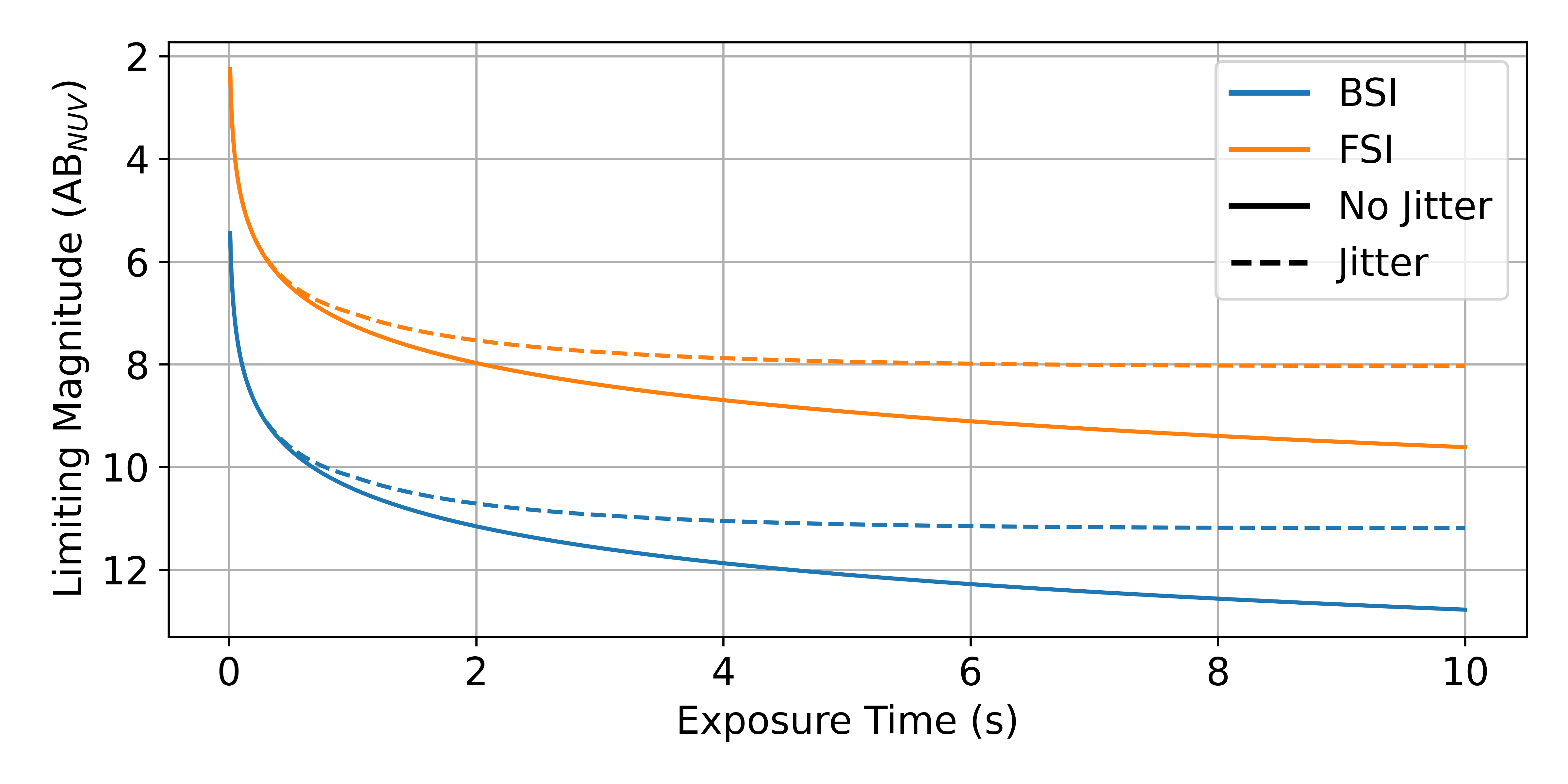}}
    \caption[Sensitivity of the LUVCam Tech Demo.]{\textit{Left:} The effective area of the complete LUVCam Tech Demo imaging system, with the back-side and front-side illuminated sensors. For the FSI sensor the QE below 300 nm has not yet been measured, here we assume the worst case of 0 QE below 300 nm. \textit{Right:} The SNR 5 isolated point source limiting magnitude vs exposure time.}
    \label{fig:performance}
\end{figure}

To predict the imaging performance of the system on-orbit, we first compute the effective area, including the filter transmission, total throughput of the optical chain, and sensor quantum efficiency. For the LUVCam tech demo flight on GRBBeta, the FSI sensor was integrated in the camera. The effective area using both the BSI and FSI versions of the sensor are shown in Fig. \ref{fig:performance}, left. Then we use the TD1 catalog and HST/STIS measured background spectrum in LEO as input, and simulate the imaging performance onto the focal plane. We include the effects of spacecraft jitter by integrating the expected ADCS jitter PSD at different exposure times to estimate the Relative Position Error (RPE).\footnote[1]{The stability during eclipse (without sun-sensor input) is not yet known, so here we use the jitter PSD expected during sun-lit operations.} Finally we simulate the sensor with as-measured read noise, dark current, and gain to produce simulated images and stellar measurements. These simulations show that a large number of sources will be accessible to LUVCam imaging under both sensor choices, and with significant margin in the ADCS jitter performance (still uncertain). This imaging capability will allow us to calibrate and track the camera performance in the orbital environment over time including total throughput (sensor QE not independently distinguishable from optics transmission), linearity, gain, saturation, and the degradation of read noise and dark current due to radiation damage.

\begin{figure}[ht]
    \centering
    \subfloat{\includegraphics[width=0.5\linewidth]{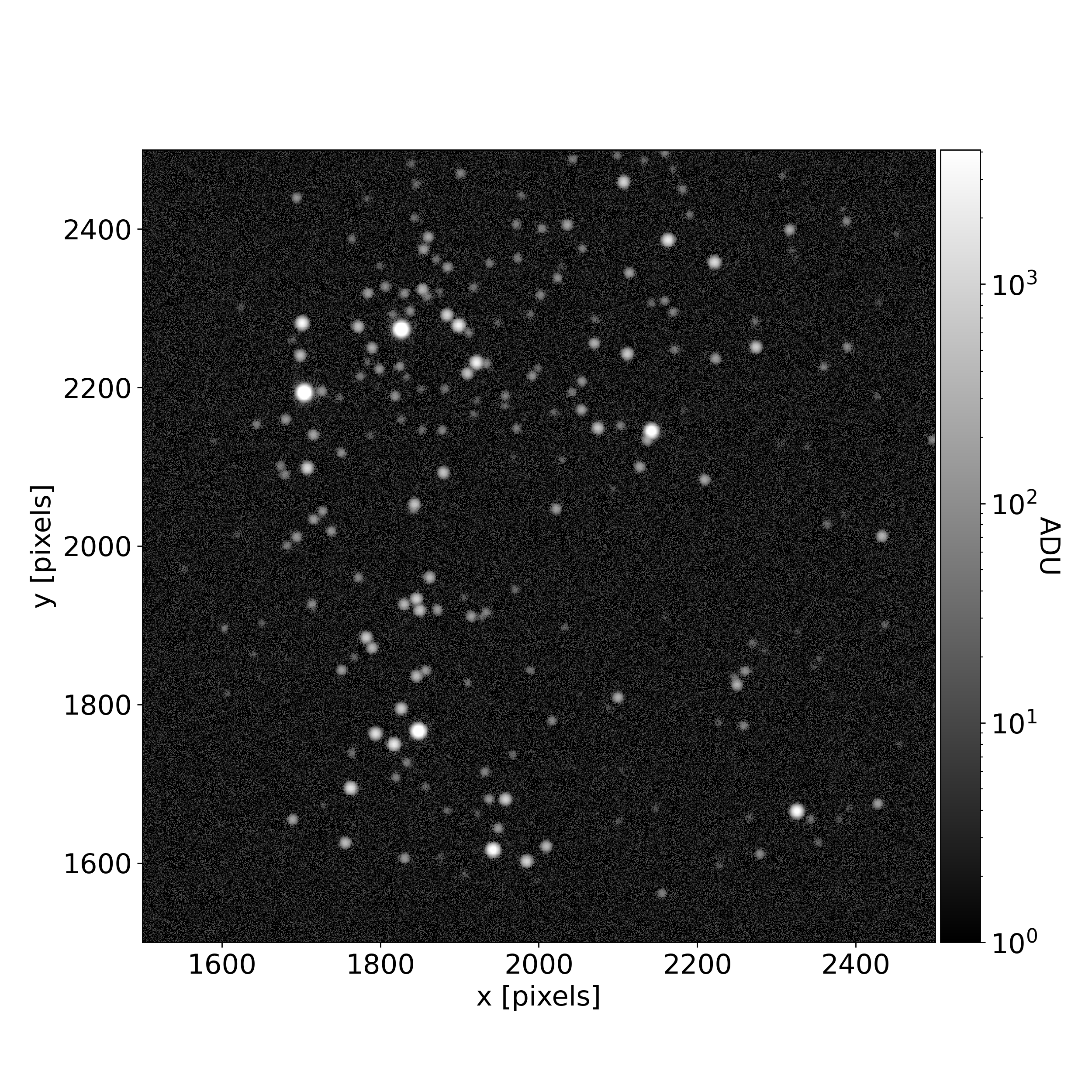}}
    \subfloat{\includegraphics[width=0.5\linewidth]{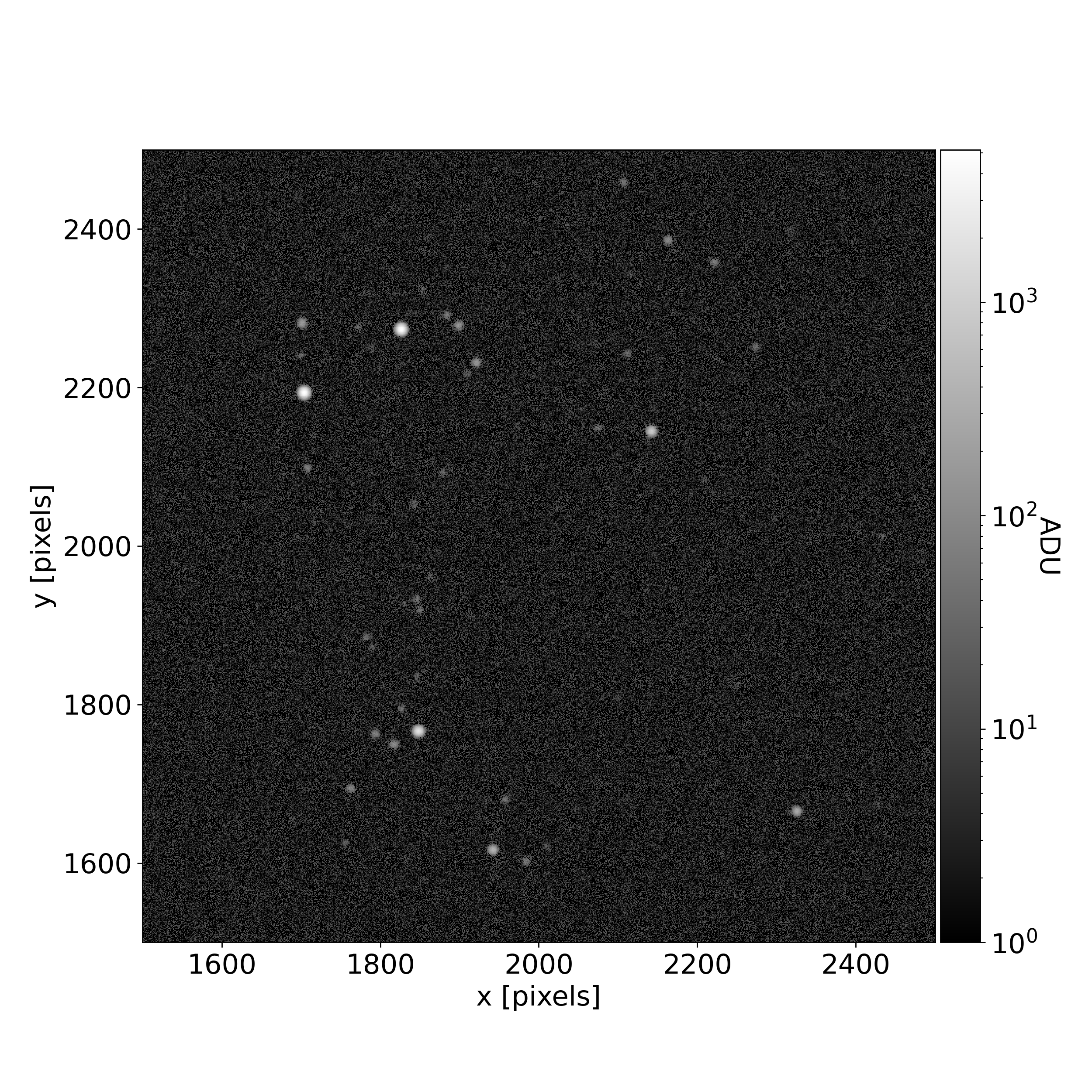}}
    \caption[Simulated image from LUVCam Tech Demo of the Orion Nebula.]{Simulated LUVCam image of the Orion Nebula (point sources only) with 1 second exposure, with the BSI (left) and FSI (right) sensors in high gain mode. The image is 9.5 degrees across. Saturation is evident in the BSI image in several sources, as is the inflated PSF from spacecraft jitter.}
    \label{fig:enter-label}
\end{figure}

\section{Conclusion}
 The LUVCam project is intended to provide a high performance, low cost, UV/optical (200-1000 nm) camera system to enable qualitatively new opportunities in space-based astronomy. The first LUVCam flight included the design, fabrication and delivery of the electronics, firmware, telescope, and thermo-mechanical structure of an orbital technology demonstration in less than 1 year. This was the Dunlap Institute's first CMOS camera development, and first orbital flight project. Many obstacles were overcome, including supply chain issues, launch schedule uncertainty, and first-timer mistakes. LUVCam on GRBBeta launched to orbit on July 9, 2024. Since then, it has been successfully commissioned. It is now undergoing characterization. The results of the characterization will be published in another paper. 
 
 The LUVCam controller is designed as a general purpose CMOS platform, and future LUVCam iterations are planned with other GPixel, Sony, and Fairchild Imaging sensors.  A second technology demonstration orbital flight, with the GSENSE4040BSI sensor, upgraded control electronics and thermal system, and extension into the far-UV, is planned for Q4 2027. The controller board architecture remains the same while there are some modifications on the FPGA and a new sensor board to accommodate the BSI sensor. Further development to support the QUVIK mission (2029 launch) is underway and will involve more precise calibration and characterization, new readout modes to support fast ROI and on-sensor guidance, and stable closed-loop thermal control.
 
\section*{Disclosures}
The authors declare there are no financial interests, commercial affiliations, or other potential
conflicts of interest that have influenced the objectivity of this research or the writing of this paper.

\section*{Code and Data Availability}
Due to the large amount of data generated for the sensor characterization, only sparse samplings were saved. These data are available from the authors upon request. Otherwise, all data in support of the findings of this paper are available within the article. 

\acknowledgments 
This project would not have been possible without the continued support of the Dunlap Institute for Astronomy \& Astrophysics, who provided our funding. The Dunlap Institute is funded through an endowment established by the David Dunlap family and the University of Toronto. We are delighted to thank many individuals who provided critical assistance along the way, for which we are extremely grateful. These include the Spacemanic team (Samuel Toman, Jan Hudec, Maksim Rezenov, Marcel Frajt, Natalia Gogolova, Petr Moravec, Jakub Kapus), Stefan Mochnacki, Corey Duce, Andy Xia, Eric LaForest, Stephanie Juneau, Max Aalto, and Roberto Abraham. NW, FM, and JR thank the support by the Czech Science Foundation (GAČR) project No. 24-11487J. 
\bibliography{report} 
\bibliographystyle{spiebib} 

\end{document}